\documentclass{article}

\usepackage[english]{babel}

\usepackage[letterpaper,top=2cm,bottom=2cm,left=3cm,right=3cm,marginparwidth=1.75cm]{geometry}

\usepackage{amsmath,caption,indentfirst}
\usepackage{graphicx}
\usepackage[colorlinks=true, allcolors=blue]{hyperref}
\usepackage{amsmath}
\title{Multidomain Model for Optic Nerve Potassium Clearance: Roles of Glial Cells and Perivascular Spaces}
\author{Shanfeng Xiao \thanks{School of Mathematical Sciences, Soochow University, 1 Shi-Zi Street, Suzhou 215006, Jiangsu Province, China}
\and Huaxiong Huang \thanks{Guangdong Provincial Key Laboratory of Interdisciplinary Research and Application for Data Science, BNU-HKBU United International College, Zhuhai 519088, China; Laboratory of Mathematics and Complex Systems, MOE, Beijing Normal University,  Beijing 100875, China;
Department of Mathematics and Statistics York University, Toronto, ON M3J 1P3, Canada}
\and Robert  Eisenberg \thanks{Department of Applied Mathematics, Illinois Institute of Technology, Chicago, IL 60616, USA}
\and  Zilong Song \thanks{Math and Statistics Department, Utah State University, Old Main Hill, Logan, UT 84322, USA}
\and Shixin Xu \thanks{Zu Chongzhi Center for Mathematics and Computational Sciences, Duke Kunshan University, 8 Duke Ave, Kunshan, China}
}

\begin{document}
	\maketitle

	\begin{abstract}
		The accumulation of potassium in the extracellular space surrounding nerve cells is a fundamental aspect of biophysics that has garnered significant attention in recent research. This phenomenon holds implications for various neurological conditions, including spreading depression, migraine, certain types of epilepsy, and potentially, learning processes. A quantitative analysis is essential for understanding the dynamics of potassium clearance following a series of action potentials. This clearance process involves multiple structures along the nerve, including glia, the extracellular space, axons, and the perivascular space, necessitating a spatially distributed systems approach akin to the cable equations of nerve physiology. In this study, we propose a multi-domain model for the optic nerve to investigate potassium accumulation and clearance dynamics. The model accounts for the convection, diffusion, and electrical migration of fluid and ions, revealing the significant roles of glia and the perivascular space in potassium buffering. Specifically, our findings suggest that potassium clearance primarily occurs through convective flow within the syncytia of glia, driven by osmotic pressure differences. Additionally, the perivascular space serves as a crucial pathway for potassium buffering and fluid circulation, further contributing to the overall clearance process. Importantly, our model's adaptability allows for its application to diverse structures with distinct channel and transporter distributions across the six compartments, extending its utility beyond the optic nerve.
	\end{abstract}

\section{Introduction}

The glymphatic system, which facilitates the clearance of metabolic waste from the central nervous system (CNS), plays a crucial role in maintaining neural health \cite{2012AParavascular,2015TheGlymphatic,2016Glymphatic,2017Impairment,2020Glymphatic,2024Modeling}. In particular, the flow of cerebrospinal fluid (CSF) through the perivascular spaces (PVS) of the brain and optic nerve is essential for clearing waste products like amyloid-$\beta$, which have been linked to neurodegenerative diseases such as Alzheimer's and cerebral amyloid angiopathy \cite{2012AParavascular,2018Flow,2022Perivascular}. Recent studies have demonstrated that CSF enters the brain along arterial PVS and drives the clearance of solutes from the interstitial fluid (ISF) at downstream locations \cite{1992Alzheimer,2016Theamyloid,verheggen_interaction_2018}. This glymphatic mechanism has also been implicated in the clearance of potassium, a process critical for maintaining ionic homeostasis in the CNS \cite{2020Glymphatic}.

The optic nerve, part of the CNS, is structurally similar to the brain in that it is surrounded by glial cells and features narrow extracellular spaces \cite{2020Quantitative}. It consists of four primary regions: the intraocular nerve head, intraorbital region, intracanalicular region, and intracranial region \cite{2009Ischemic}. This study focuses on the intraorbital region, which constitutes the majority of the optic nerve. The optic nerve is crucial in waste clearance, with CSF entering via perivascular spaces around blood vessels and interacting with the glial cells that line these spaces \cite{2017Evidence}. Understanding the dynamics of fluid flow and ionic transport in the optic nerve is essential, especially in the context of diseases like glaucoma, which are closely associated with impaired fluid regulation and waste clearance \cite{2017Aquaporin}.

Glial cells, particularly astrocytes, play a significant role in potassium clearance in the optic nerve. During neural activity, potassium ions accumulate in the extracellular space, and their efficient clearance is necessary to maintain normal neuronal function and prevent excitotoxicity. Glial cells form an interconnected network via connexin-based gap junctions, which allows them to act as a syncytium, facilitating the redistribution and clearance of potassium from the extracellular space \cite{2017Astrocytic}. This process is also closely linked to fluid flow within the glymphatic system, as glial cells regulate fluid movement through aquaporin-4 (AQP4) channels located in their endfeet \cite{2006Theimpact,2011Anaquaporin4}.

Mathematical models have been developed to describe the behavior of the optic nerve, primarily focusing on fluid dynamics and ionic transport. Early models typically considered fluid flow within the perivascular space, driven by arterial pulsations and volume changes \cite{2016Glymphatic,2003Arterial}. More recently, machine learning techniques have also been applied to explore the mechanisms within the perivascular spaces \cite{boster2023artificial,chou2024machine}, with a detailed review available in \cite{bohr2022glymphatic}. 

However, most of these models do not account for the interactions between different compartments, such as the extracellular space, glial compartments, and perivascular spaces. Additionally, the coupling between fluid flow, electric fields, and mass transport is often neglected. In this study, we build on previous work by Zhu et al. \cite{2020ATridomain} and introduce a multidomain model that incorporates these compartments, capturing the full complexity of ionic electrodiffusion, osmosis, and fluid circulation in the optic nerve. Our model also integrates key aspects of glial function and perivascular space dynamics, particularly their role in potassium clearance through convective, diffusive, and electrochemical transport mechanisms \cite{2019ABidomain}.

Our model also considers the interaction between the optic nerve and its surrounding CSF, which flows through the perivascular spaces and subarachnoid space (SAS). This fluid serves as both a source of nutrients and a pathway for waste clearance \cite{2017Mechanisms}. By differentiating between the extracellular and perivascular spaces, and by including the direct communication between CSF and perivascular spaces A and V, we aim to provide a more detailed understanding of the optic nerve's glymphatic function, particularly in response to neural activity and potassium clearance.  

The remainder of this article is organized as follows. Section 2 presents the detailed mathematical model for microcirculation in the optic nerve. In Section 3, we provide the simulation results and discuss the effects of glial cells and perivascular spaces on potassium clearance. Finally, concluding remarks are provided in Section 4.

	\begin{figure}
		\centering
		\includegraphics[width=0.8\linewidth]{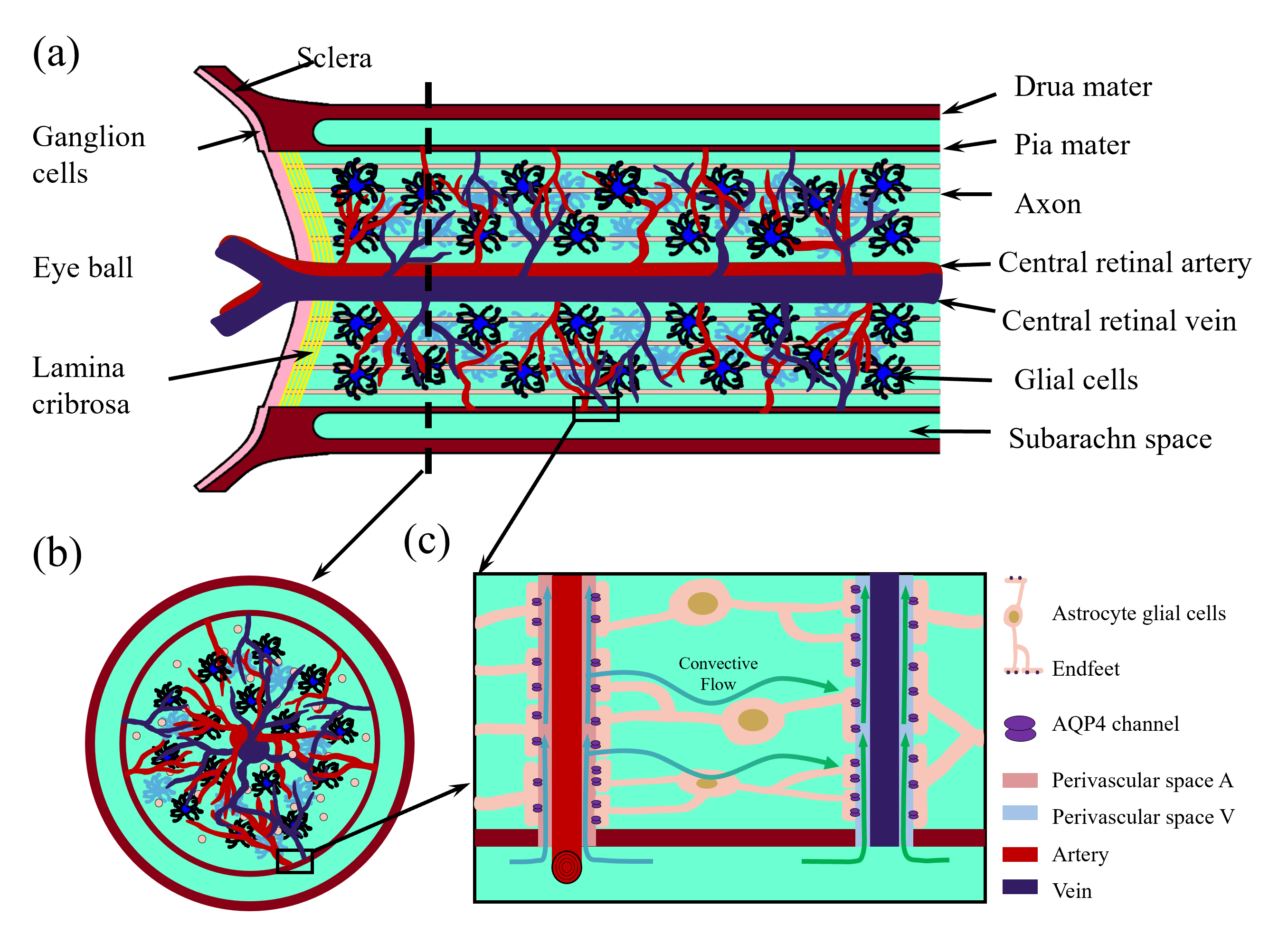}
		\caption{\label{fig:Optic-nerve}Optic nerve structure. (a) Longitudinal section of the optic nerve; (b) cross section of the optic nerve.}
	\end{figure}

	\section{Mathematical Model of Microcirculation of the Optic Nerve}
	Fig. \ref{fig:Optic-nerve} shows the structure we are modeling. This is a subset of all the structures of interest in the optic nerve, but it is a good place to start. Computational domain $\Omega$  consists of the subarachnoid space (SAS) region $\Omega_{SAS}$ and optic nerve region $\Omega_{OP}$, i.e.	
	$$\Omega=\Omega_{OP}\cup\Omega_{SAS},~~ \Omega_{OP}\cap\Omega_{SAS} =\Gamma_7,$$
	where the domain $\Omega_{SAS}$ is filled with CSF, enclosed by dura mater $\Gamma_7$ and pia mater $\Gamma_4$. 
	
	
	For the optic nerve region, based on the model first proposed in Ref. \cite{2020ATridomain}, we introduce a six-domain model for microcirculation of the optic nerve, which includes blood vessels. We must remind the reader that the optic nerve bundler contains many nerve fibers, at lease one blood vessel, and a glial syncytium as shown in Fig. \ref{fig:Optic-nerve}. In addition to the axon ($\Omega_{ax}$), glial ($\Omega_{gl}$), and extracellular ($\Omega_{ex}$) compartments, the model incorporates three perivascular spaces: perivascular space A (surrounding the artery, $\Omega_{pa}$), perivascular space V (surrounding the vein, $\Omega_{pv}$), and perivascular space C (surrounding the capillary, $\Omega_{pc}$).
		$$\Omega_{OP}=\Omega_{gl}\cup\Omega_{ax}\cup\Omega_{ex}\cup\Omega_{pa}\cup\Omega_{pv}\cup\Omega_{pc}.$$
	
	For the boundaries, $\Gamma_1$ is the central retinal blood wall of the optic nerve; $\Gamma_2$ and $\Gamma_3$ are the far end (away from the eyeball) of the optic nerve which is connected to optic canal region \cite{1984Thesheath}. $\Gamma_5$ is used to model the dura mater connected to the sclera (the white matter of the eye) and assumed to be non-permeable \cite{2009Ischemic}. $\Gamma_6$ is used to denote the lamina cribrosa where the optic nerve head exits the eye posteriorly through pores of the lamina cribrosa \cite{2003Anatomic}. 
	
	

	
	\begin{table}
		\centering
		\caption*{\textbf{GLOSSARY}}
		\scalebox{0.7}{
			\begin{tabular}{ll}
				$C_{l}^{i}$: Ion $i$ concentration in compartment $l$, & $k_{e}^{l}$: Electroosmotic in compartment $l$, \\
				$\phi_{l}$: Electric potential in compartment $l$, & $\mathcal{M}_{a,b}$: Interface area form compartment $a$ to $b$ in per unit control volume, \\
				$p_{l}$: Hydrostatic pressure in compartment $l$, & $\kappa_{l}$: fluid permeability in compartment $l$, \\
				$\eta_{l}$:  Volume fraction in compartment $l$, & $\mu$: Fluid viscosity, \\
				$O_{l}$: Osmotic concentration in compartment $l$, & $L_{a,b}$: Hydrostatic permeability of interface form $a$ to $b$, \\
				$\mathbf{u}_{l}$: Fluid velocity inside of compartment $l$, & $g_{a,b}^{i}$: Conductance of interface form compartment $a$ to $b$ for ion $i$, \\
				$\mathbf{j}_{l}$: Ion $i$ flux inside of compartment $l$, & $\bar{g}^{i}$: Maximum conductance of axon membrane for ion $i$, \\
				$U_{a,b}$: Fluid flux across the interface form  compartment $a$ to $b$, & $g_{leak}^{i}$: Leak conductance of axon membrane for ion $i$, \\
				$J_{a,b}^{i}$: Ion $i$ flux across the interface form compartment $a$ to $b$, & $K_{k}$: Stiffness constant of the interface for compartment $k$, \\
				$J_{a,b}^{p,i}$: Active ATP based ion $i$ pump form compartment $a$ to $b$, & $\tau_{l}$: Tortuosity of compartment $l$, \\
				$J_{a,b}^{c,i}$: Passive source form compartment $a$ to $b$, & $D_{l}^{i}$: Diffusion coefficient of $i$ ion in compartment $l$, \\
				$I_{a,1}$: Max current of $\alpha_{1}-Na/k$ pump on $a$ membrane, & $T$: Temperature, \\
				$I_{a,2}$: Max current of $\alpha_{2}-Na/k$ pump on $a$ membrane, & $k_{B}$: Boltzmann constant, \\
				$A_{l}$: Negative charged protein density in compartment $l$, & $e$: The magnitude of the electron charge,\\
				$N_{A}$: Avogadro constant. &  
			\end{tabular}
		}
	\end{table}
	
	\begin{figure}
		\centering
		\includegraphics[width=0.7\linewidth]{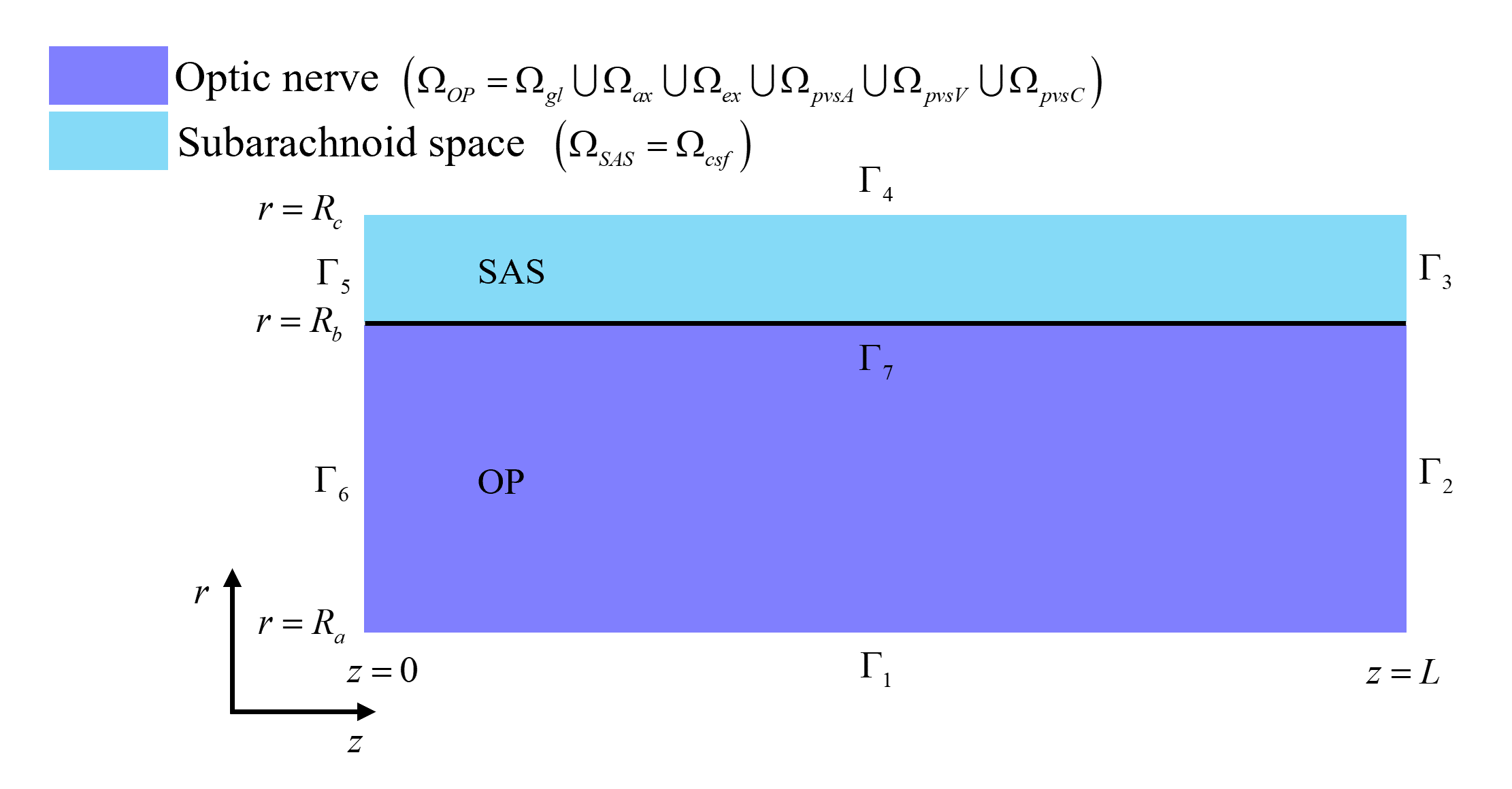}
  \caption{\label{fig:Op_structure_model}The optic nerve $\Omega_{OP}$ consists of axon compartment, glial compartment, extracellular space, perivascular space. The subarachnoid space $\Omega_{SAS}$ only has cerebrospinal fluid.}
	\end{figure}
	
	
	
	The model is mainly derived from laws of conservation of ions and fluid for flow through membranes between intracellular compartments and extracellular space (ECS) \cite{2001Diffusion}, in $\Omega_{l}, l=ax, gl, ex, pa, pv, pc$
	
	\begin{equation}
		\dfrac{\partial}{\partial t}(\eta_lf_l)+\nabla\cdot(\eta_l\mathbf{J}_l)+S=0,
	\end{equation}
	
	\noindent where $\eta_l$ is the volume fraction of $l$ compartment, $f_l$ is the concentration of a given substance, $\mathbf{J}_l$ is the flux inside the compartment, and $S$ is the source term induced by trans-membrane communications due to the active pumps and passive leak channels on the membranes.
	

	
\begin{figure}
		\centering
		\includegraphics[width=0.7\linewidth]{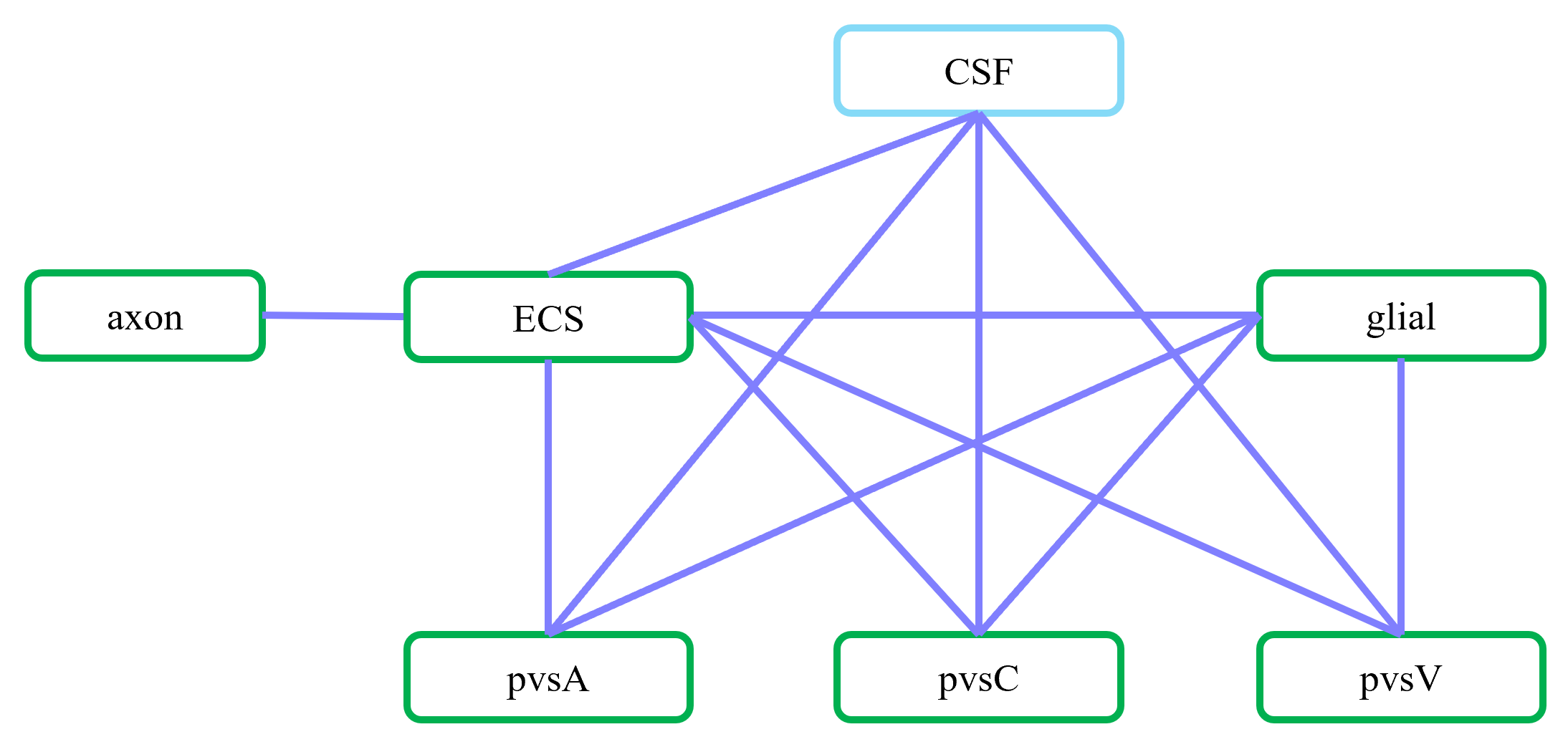}
		\caption{\label{fig:Model-exchange}
			In the optic nerve $\Omega_{OP}$ region, the ECS exchanges fluid and ions with the axon, glial, the pvsA, the pvsV and the pvsC; glial exchanges fluid and ion with the ECS, the pvsA, the pvsV and the pvsC; CSF exchanges fluid and ions with the ECS, the pvsA, the pvsV, and the pvsC by across the pia mater in $\Gamma_{7}$, see Fig. \ref{fig:Op_structure_model}.}
	\end{figure}

	Based on the structure of the optic nerve, we have the following global assumptions for the model:
	\begin{itemize}
		\item \textbf{Axial symmetry:} For simplicity, axial symmetry is assumed. The model can be straightforwardly extended to three dimensions when data and needs justify the considerable extra computational resources needed to analyze such models.
		
		\item\textbf{Ions type:} For simplicity, only three main types of ions, $\rm{Na^+}, \rm{K^+}, \rm{Cl^-}$ are taken into consideration. 
		\item\textbf{Charge neutrality:} In each domain, we assume electroneutrality \cite{1999Transport}
		\begin{subequations}\label{Chargeneutrality}
			\begin{align}
				& \eta_{gl}\sum_iz^iC_{gl}^i+z^{gl}\eta_{gl}^{re}A_{gl}=0,\\
				& \eta_{ax}\sum_iz^iC_{ax}^i+z^{ax}\eta_{ax}^{re}A_{ax}=0,\\
				& \sum_iz^iC_{l}^i=0, l=ex, pa, pv, pc,csf,
			\end{align}
		\end{subequations}
		where $A_{l}>0$ with $l=ax, gl$ is the  density of proteins in axons or glial cells. The proteins are negatively charged, but the charge density is customarily described by a positive number. We assume the permanent negatively charged protein is uniformly distributed within glial cells and axons at resting state and has valence $z_{l}^{-1}, l=ax, gl$. The $\eta_{ax}$ and $\eta_{gl}$ are the volume fraction of axon and glial compartments in the optic nerve and $\eta_{ax}^{re}$ and $\eta_{gl}^{re}$ are the resting state volume fractions.
		\item  \textbf{Anisotropy of axon compartment and isotropy of other compartments:} The axons are separated, more or less parallel cylindrical cells that form separate compartments. They are electrically isolated and molecules cannot diffuse directly from one to another. So the intra-compartment fluxex of ions and fluid are only along the axis direction.  In contrast, other compartments are fully connected. For example, the glial cells are connected by connexins and form a syncytium.  So the intra-compartment fluxes flow in both axis direction and radius direction. 
		
		\item \noindent \textbf{Communications between compartments:} The communications among different compartments are illustrated in Fig.\ref{fig:Model-exchange}. Especially,  there is no direct interaction between glial (or perivascular space) and axon compartments.  Interactions occur only through changes in concentration, electrical potential, and flows in ECS \cite{Sibille2015The}.
		
		
		
		
		
	\end{itemize}

	\subsection{Fluid Circulation}
	In this subsection, we present the fluid circulation model.  First, due to the conservation law, the volume fraction of each compartment $\eta_l, l= ax, gl,ex, pa,pv,pc$ satisfy
	
	\begin{subequations}
		\begin{align}
			& \dfrac{\partial \eta_{ax}}{\partial t}+\mathcal{M}_{ax,ex}U_{ax,ex}+\dfrac{\partial}{\partial z}(\eta_{ax}u_{ax}^{z})=0, \\
			\begin{split}
				& \dfrac{\partial \eta_{gl}}{\partial t}+\mathcal{M}_{gl,ex}U_{gl,ex}+\mathcal{M}_{gl,pa}U_{gl,pa}+\mathcal{M}_{gl,pv}U_{gl,pv}+\mathcal{M}_{gl,pc}U_{gl,pc}+\nabla\cdot(\eta_{gl}\mathbf{u}_{gl})=0, \\
			\end{split}\\
			\begin{split}
				& \dfrac{\partial \eta_{pa}}{\partial t}+\mathcal{M}_{pa,ex}U_{pa,ex}-\mathcal{M}_{gl,pa}U_{gl,pa}+\mathcal{M}_{pa,pc}U_{pa,pc}+\nabla\cdot(\eta_{pa}\mathbf{u}_{pa})=0,\\
			\end{split}\\
			\begin{split}
				& \dfrac{\partial \eta_{pv}}{\partial t}+\mathcal{M}_{pv,ex}U_{pv,ex}-\mathcal{M}_{gl,pv}U_{gl,pv}-\mathcal{M}_{pc,pv}U_{pc,pv}+\nabla\cdot(\eta_{pv}\mathbf{u}_{pv})=0,\\
			\end{split}\\
			\begin{split}
				& \dfrac{\partial \eta_{pc}}{\partial t}+\mathcal{M}_{pc,ex}U_{pc,ex}-\mathcal{M}_{gl,pc}U_{gl,pc}-\mathcal{M}_{pa,pc}U_{pa,pc}+\mathcal{M}_{pc,pv}U_{pc,pv}+\nabla\cdot(\eta_{pc}\mathbf{u}_{pc})=0,\\
			\end{split} \\
			\begin{split}
				& \dfrac{\partial}{\partial z}(\eta_{ax}u_{ax}^{z}) + \sum_{k=gl,ex,pa,pv,pc}\nabla\cdot(\eta_{k}\mathbf{u}_{k})=0,\\
			\end{split}\\
			& \eta_{ax}+\eta_{gl}+\eta_{ex}+\eta_{pa}+\eta_{pv}+\eta_{pc}=1,
		\end{align}
	\end{subequations}
	where $U_{l,k}$ is the fluid velocity across the membrane/interface between $l_{th}$ and $k_{th}$ compartments with  surface volume ratio $\mathcal{M}_{l,k}$ and $\mathbf{u}_l$ is the fluid velocity inside the $l_{th}$ compartment.

	
	
	
	
	The transmembrane fluid flux is proportional to the intracellular/extracellular hydrostatic pressure and osmotic pressure differences, i.e., Starling's law on the membrane, while the fluid flow from perivascular space A to perivascular space C and from perivascular space C to perivascular space V are only proportional to the difference of hydrostatic pressure due to the direct connection.
	
	\begin{subequations}
		\begin{align}
			& U_{l,ex}=L_{l,ex}(p_{l}-p_{ex}-\gamma_{l,ex}RT(O_{l}-O_{ex})),~l=ax,gl,pa,pv,pc \\
			& U_{gl,l}=L_{gl,l}(p_{gl}-p_{l}-\gamma_{gl,l}RT(O_{gl}-O_{l})),~l=pa,pv,pc. \\
		\end{align}
	\end{subequations}
	Here  $\gamma_{l,k}$  and $L_{l,k}$ are the reflection coefficient \cite{2017Quantitative} and hydraulic permeability of the membrane between $l_{th}$ and $k_{th}$ compartments, respectively. In this paper osmotic pressure is defined by $RTO_l$ \cite{2019ABidomain,2018Osmosis}  
	$$O_{l}=\sum_i C_{l}^i+A_{l}\dfrac{\eta_{l}^{re}}{\eta_{l}},  \quad  l=ax, gl,$$
	$$O_{l}=\sum_i C_{l}^i, \quad  l=ex, pa, pv, pc,$$
	where $A_{l}\dfrac{\eta_{l}^{re}}{\eta_{l}}>0$ is the density of the permanent negatively charged protein in glial cells and axons that varies with the volume (fraction) of the region,  $R$ is the molar gas constant, and $T$ is temperature.
	
	The hydrostatic pressure $p_l$ and the volume fraction $\eta_{l}$ are connected by the force balance on the membrane \cite{2018Osmosis,Mori2015A}. The membrane force is balanced by the hydrostatic pressure difference on both sides of the semipermeable membrane. Then the variation of volume fraction from the resting state is proportional to the variation of hydrostatic pressure difference from the resting state.
	
	\begin{subequations}\label{eq:volumefraction}
		\begin{align}
			& K_{l}(\eta_{l}-\eta_{l}^{re})=p_{l}-p_{ex}-(p_{l}^{re}-p_{ex}^{re}),
		\end{align}
	\end{subequations}
	
	\noindent where $K_{l}$ is the stiffness constant and $\eta_{l}^{re}$ and $p_l^{re}$ are the resting state volume fraction and hydrostatic pressure of $l_{th}$ compartment with $l=ax,gl,pa,pv,pc$.
	
	The subarachnoid space region is modeled as a porous media filled with CSF. Therefore, the solution is incompressible in the $\Omega_{SAS}$, and we have 
	
	\begin{equation}
		\nabla\cdot\mathbf{u}_{csf}=0, \quad in \ \Omega_{SAS}.
	\end{equation}
	
	In the next section, we provide submodels of fluid velocities inside each compartment and the corresponding boundary conditions. 

    In this paper, we define a membrane boundary condition to describe the fluid or ion communication in the connect interface between the optic nerve and the retina or the optic canal or orbital. We assume this fluid or ion communication depend on the difference in pressure or ion concentration between the two sides of this connect interface $U_{int}=L(P_{in}-P_{out})$. Where $U_{int}$ is the interface velocity, $P_{in}$ and $P_{out}$ are hydrostatic pressure on both sides of the interface, respectively.
 
	\textbf{Fluid Velocity in the Glial Compartment.} The glial space is a connected space, where the intracellular fluid can flow from cell to cell through connexin proteins joining membranes of neighboring cells. The velocity of fluid in glial syncytium $\mathbf{u}_{gl}$ depends on the gradients of hydrostatic pressure and osmotic pressure \cite{2019ABidomain,2018Osmosis,Mori2015A,1985Steady,1979Electrical} as 
	
	\begin{subequations}
		\begin{align}
			u_{gl}^{r}& =-\dfrac{\kappa_{gl}\tau_{gl}}{\mu}\left(\dfrac{\partial p_{gl}}{\partial r}-\gamma_{gl}RT\dfrac{\partial O_{gl}}{\partial r}\right),\\
			u_{gl}^{\theta}& =-\dfrac{\kappa_{gl}\tau_{gl}}{\mu}\left(\dfrac{1}{r}\dfrac{\partial p_{gl}}{\partial \theta}-\gamma_{gl}RT\dfrac{1}{r}\dfrac{\partial O_{gl}}{\partial \theta}\right),\\
			u_{gl}^{z}& =-\dfrac{\kappa_{gl}\tau_{gl}}{\mu}\left(\dfrac{\partial p_{gl}}{\partial z}-\gamma_{gl}RT\dfrac{\partial O_{gl}}{\partial z}\right),
		\end{align}
	\end{subequations}
	where $\kappa_{gl}$ and $\tau_{gl}$  are the permeability and tortuosity of the glial compartment.

	For the boundary condition, on the left and right boundaries of domain $\Omega_{OP}$, the membrane boundary condition is used since they connect to intraocular and intracanalicular regions. On the top and bottom boundary, the no-flux boundary condition is used 
	
	\begin{equation}
		\begin{cases}
			\mathbf{u}_{gl}\cdot\hat{\mathbf{n}}_{r}=0, & \mbox{on} \ \Gamma_1\\
			\mathbf{u}_{gl}\cdot\hat{\mathbf{n}}_{z}=L_{gl,right}(p_{gl}-p_{gl,right}),& \mbox{on} \ \Gamma_2\\
			\mathbf{u}_{gl}\cdot\hat{\mathbf{n}}_{z}=L_{gl,left}(p_{gl}-p_{gl,left}), & \mbox{on} \ \Gamma_6\\
			\mathbf{u}_{gl}\cdot\hat{\mathbf{n}}_{r}=0, & \mbox{on} \ \Gamma_7\\
		\end{cases}
	\end{equation}
	where 	$\hat{\mathbf{n}}_r$ and $\hat{\mathbf{n}}_{z}$ are the outward normal vector of domain $\Omega_p$.
	
	\textbf{Fluid Velocity in the Axon Compartment.}  Since the axons are only connected in the longitudinal direction, the fluid velocity in the region of the axon is defined  as
	
	\begin{subequations}
		\begin{align}
			u_{ax}^{r}& =0,\\
			u_{ax}^{\theta}& =0,\\
			u_{ax}^{z}& =-\dfrac{\kappa_{ax}}{\mu}\dfrac{\partial p_{ax}}{\partial z},
		\end{align}
	\end{subequations}
	where $\kappa_{ax}$ is the permeability  of the axon compartment. 
	
	Similarly, membrane boundary conditions are used on the left and right boundaries 
	
	\begin{equation}
		\begin{cases}
			\mathbf{u}_{ax}\cdot\hat{\mathbf{n}}_{z}=L_{ax,right}(p_{ax}-p_{ax,right}), & \mbox{on} \ \Gamma_2\\
			\mathbf{u}_{ax}\cdot\hat{\mathbf{n}}_{z}=L_{ax,left}(p_{ax}-p_{ax,left}), & \mbox{on} \ \Gamma_6\\
		\end{cases}
	\end{equation}
	
	\textbf{Fluid Velocity in the Extracellular and Perivascular Spaces}  Since both the extracellular and perivascular spaces  are narrow, the  velocity is determined by the gradients of hydro-static pressure and electric potential \cite{2019ABidomain,2014Self,2012Development}, for $l=ex,pa,pv,pc$ 
	\begin{subequations}
		\begin{align}
			u_{l}^{r}& =-\dfrac{\kappa_{l}\tau_{l}}{\mu}\dfrac{\partial p_{l}}{\partial r}-k_e^{l}\tau_{l}\dfrac{\partial \phi_{l}}{\partial r},\\
			u_{l}^{\theta}& =-\dfrac{\kappa_{l}\tau_{l}}{\mu}\dfrac{1}{r}\dfrac{\partial p_{l}}{\partial \theta}-k_e^{l}\tau_{l}\frac{1}{r}\dfrac{\partial \phi_{l}}{\partial \theta},\\
			u_{l}^{z}& =-\dfrac{\kappa_{l}\tau_{l}}{\mu}\dfrac{\partial p_{l}}{\partial z}-k_e^{l}\tau_{l}\dfrac{\partial \phi_{l}}{\partial z},
		\end{align}
	\end{subequations}
	
	where $\phi_{l}$ is the electric potential, $\tau_{l}$ is the tortuosity \cite{2001Diffusion,1995Extracellular}, $k_e^l$ describes the effect of electroosmotic flow \cite{2014Self,2012Development,1985Electro}, $\kappa_{l}$ is the permeability. 
	
	For the boundary conditions, due to the connections to  intraocular region and intracanalicular region on the left and right boundaries, respectively,  membrane boundary conditions are used for extracellular and perivascular spaces A \& V. Nonflux boundary condition is used for perivascular space c on $\Gamma_2$ and $\Gamma_6$. On $\Gamma_1$, due to the central blood vessels, Dirichlet boundary conditions on pressure are used for perivascular space A \& V.  For the ECS and pVSC, the zero penetration velocity is used. On the pia mater $\Gamma_7$, the CSF could directly communicate with the perivascular spaces A \& V due to the hydrostatic pressure difference. However, for ECS and pvsc, the CSF leaks into these two compartments through membranes. Then the velocity is fixed as the transmembrane velocity which depends on hydrostatic and osmotic pressure difference.  In summary, the conditions are listed as follows

	\begin{equation}
		\begin{cases}
			p_{pa}=p_{PA},~ p_{pv}=p_{PV}, ~\mathbf{u}_{pc}\cdot\hat{\mathbf{n}}_{r}=0, ~\mathbf{u}_{ex}\cdot\hat{\mathbf{n}}_{r}=0,& \mbox{on} \ \Gamma_1\\
			\mathbf{u}_{pa}\cdot\hat{\mathbf{n}}_{z}=L_{pa,right}(p_{pa}-p_{pa,right}), \mathbf{u}_{pv}\cdot\hat{\mathbf{n}}_{z}=L_{pv,right}(p_{pv}-p_{pv,right}),   & \mbox{on}  \ \Gamma_2\\
			\mathbf{u}_{pc}\cdot\hat{\mathbf{n}}_{z}=0,\mathbf{u}_{ex}\cdot\hat{\mathbf{n}}_{z}=L_{ex,right}(p_{ex}-p_{pv,right}), & \mbox{on}  \ \Gamma_2\\
			\mathbf{u}_{pa}\cdot\hat{\mathbf{n}}_{z}=L_{pa,left}(p_{pa}-p_{pa,left}),\mathbf{u}_{pv}\cdot\hat{\mathbf{n}}_{z}=L_{pv,left}(p_{pv}-p_{IOP}), & \mbox{on}  \ \Gamma_6\\
			\mathbf{u}_{pc}\cdot\hat{\mathbf{n}}_{z}=0, \mathbf{u}_{ex}\cdot\hat{\mathbf{n}}_{z}=L_{ex,left}(p_{ex}-p_{IOP}), & \mbox{on}  \ \Gamma_6\\
			\mathbf{u}_{pa}\cdot\hat{\mathbf{n}}_{r}=L_{pia,pa}(p_{pa}-p_{csf})), 	\mathbf{u}_{pv}\cdot\hat{\mathbf{n}}_{r}=L_{pia,pv}(p_{pv}-p_{csf})), & \mbox{on}  \ \Gamma_7\\
			\mathbf{u}_{pc}\cdot\hat{\mathbf{n}}_{r}=L_{pia,pc}(p_{pc}-p_{csf}-\gamma_{pia}RT(O_{pc}-O_{csf})), &\mbox{on}  \ \Gamma_7\\
			\mathbf{u}_{ex}\cdot\hat{\mathbf{n}}_{r}=L_{pia,ex}(p_{ex}-p_{csf}-\gamma_{pia}RT(O_{ex}-O_{csf})),  &\mbox{on}  \ \Gamma_7.
		\end{cases}
	\end{equation}

	\noindent where $p_{IOP}$ is the intraocular pressure (IOP), $\gamma_{pia}$ is the  the reflection coefficient of pia mater.
	
	
	
	\textbf{Fluid Velocity in the SAS Region.} The cerebrospinal fluid velocity in the SAS region is determined by the gradients of hydro-static pressure and electric potential
	
	\begin{subequations}
		\begin{align}
			u_{csf}^{r}& =-\dfrac{\kappa_{csf}\tau_{csf}}{\mu}\dfrac{\partial p_{csf}}{\partial r}-k_e^{csf}\tau_{csf}\dfrac{\partial \phi_{csf}}{\partial r},\\
			u_{csf}^{\theta}& =-\dfrac{\kappa_{csf}\tau_{csf}}{\mu}\dfrac{1}{r}\dfrac{\partial p_{csf}}{\partial \theta}-k_e^{csf}\tau_{csf}\frac{1}{r}\dfrac{\partial \phi_{csf}}{\partial \theta},\\
			u_{csf}^{z}& =-\dfrac{\kappa_{csf}\tau_{csf}}{\mu}\dfrac{\partial p_{csf}}{\partial z}-k_e^{csf}\tau_{csf}\dfrac{\partial \phi_{csf}}{\partial z},
		\end{align}
	\end{subequations}
	
	The fluid flow across the semi-permeable membrane $\Gamma_4$ is produced by the lymphatic drainage on the dura membrane, which depends on the difference between cerebrospinal fluid pressure and orbital pressure (OBP). At boundary $\Gamma_3$, we assume the hydrostatic pressure is equal to the cerebrospinal fluid pressure. At boundary $\Gamma_5$,  a non-permeable boundary is used.  On the pia membrane $\Gamma_7$,  the total CSF transmembrane velocity is determined by the conservation law, 
	
	\begin{equation}
		\begin{cases}
			p_{csf}=p_{CSF}, & \mbox{on} \ \Gamma_3\\
			\mathbf{u}_{csf}\cdot\hat{\mathbf{n}}_{r}=L_{dr}(p_{csf}-p_{OBP}), & \mbox{on} \ \Gamma_4\\
			\mathbf{u}_{csf}\cdot\hat{\mathbf{n}}_{z}=0, & \mbox{on} \ \Gamma_5\\
			\mathbf{u}_{csf}\cdot\hat{\mathbf{n}}_{r}=L_{pia,ex}(p_{ex}-p_{csf}-\gamma_{pia}RT(O_{ex}-O_{csf}))\\
			\ \ \ \ \ \ \ \ \ \ \ \ +L_{pia,pa}(p_{pa}-p_{csf}) \\
			\ \ \ \ \ \ \ \ \ \ \ \ +L_{pia,pv}(p_{pv}-p_{csf})\\
			\ \ \ \ \ \ \ \ \ \ \ \ +L_{pia,pc}(p_{pc}-p_{csf}-\gamma_{pia}RT(O_{pc}-O_{csf})), & \mbox{on} \ \Gamma_7\\
		\end{cases}
	\end{equation}
	
	\noindent where $p_{CSF}$ is the cerebrospinal fluid pressure \cite{2009Intracellular} in boundary $\Gamma_3$.
	
	\subsection{Ion Transport}
	
	The conservation of ion species implies the following system of partial differential equations to describe the dynamics of ions in each region, for $i=\rm{Na^+},~\rm{K^+},~\rm{Cl^-}$ in domain $\Omega_{OP}$
	
	\begin{subequations}\label{eq:Iontransport}
		\begin{align}
			& \dfrac{\partial \eta_{ax}C_{ax}^{i}}{\partial t}+\mathcal{M}_{ax,ex}J_{ax,ex}^{i}+\dfrac{\partial}{\partial z}(\eta_{ax}j_{ax,z}^{i})=0,\\
			& \dfrac{\partial \eta_{gl}C_{gl}^{i}}{\partial t}+\mathcal{M}_{gl,ex}J_{gl,ex}^{i}+\mathcal{M}_{gl,pa}J_{gl,pa}^{i}+\mathcal{M}_{gl,pv}J_{gl,pv}^{i}+\mathcal{M}_{gl,pc}J_{gl,pc}^{i}+\nabla\cdot(\eta_{gl}\mathbf{j}_{gl}^{i})=0,\\
			& \dfrac{\partial \eta_{pa}C_{pa}^{i}}{\partial t}+\mathcal{M}_{pa,ex}J_{pa,ex}^{i}-\mathcal{M}_{gl,pa}J_{gl,pa}^{m,i}+\mathcal{M}_{pa,pc}J_{pa,pc}^{i}+\nabla\cdot(\eta_{pa}\mathbf{j}_{pa}^{i})=0,\\
			& \dfrac{\partial \eta_{pv}C_{pv}^{i}}{\partial t}+\mathcal{M}_{pv,ex}J_{pv,ex}^{i}-\mathcal{M}_{gl,pv}J_{gl,pv}^{m,i}-\mathcal{M}_{pc,pv}J_{pc,pv}^{i}+\nabla\cdot(\eta_{pv}\mathbf{j}_{pv}^{i})=0,\\
			& \dfrac{\partial \eta_{pc}C_{pc}^{i}}{\partial t}+\mathcal{M}_{pc,ex}J_{pc,ex}^{i}-\mathcal{M}_{gl,pc}J_{gl,pc}^{i}-\mathcal{M}_{pa,pc}J_{pa,pc}^{i}+\mathcal{M}_{pc,pv}J_{pc,pv}^{i}+\nabla\cdot(\eta_{pc}\mathbf{j}_{pc}^{i})=0,\\
			& \dfrac{\partial \eta_{ex}C_{ex}^{i}}{\partial t}-\sum_{k=ax,gl,pa,pv,pc}\mathcal{M}_{k,ex}J_{k,ex}^{i}+\nabla\cdot(\eta_{ex}\mathbf{j}_{ex}^{i})=0,
		\end{align}
	\end{subequations}
	and in the $\Omega_{SAS}$ region,
	\begin{equation}
		\dfrac{\partial C_{csf}^{i}}{\partial t}+\nabla\cdot(\mathbf{j}_{csf}^{i})=0.
	\end{equation}
	
	\paragraph{Transmembrane Ion Flux}
	The transmembrane ion flux $J_{l,k}^{i}$ ( $l,k=ax,ex;gl,ex;gl,pa;gl,pv;gl,pc;$) consists of an active ion pump source $J_{l,k}^{p,i}$ and passive ion channel source $J_{l,k}^{c,i}$,
	
	$$J_{l,k}^{i}=J_{l,k}^{p,i}+J_{l,k}^{c,i};\ \  i=\rm{Na^+},~\rm{K^+},~\rm{Cl^-}.$$
	
	Due to the gaps between astrocytes endfeet \cite{ray2021quantitative}, 	the transmembrane ion flux $J_{l,k}^{i}$ between perivascular spaces and ECS $(l,k=pa,ex;pv,ex;pc,ex)$ consists of direct transportation with fluid $C_{up,wind}^{i}U_{l,k}$ and passive ion channel source $J_{l,k}^{c,i}$,
	
	$$J_{l,k}^{i}=C_{up,wind}^{i}U_{l,k}+J_{l,k}^{c,i};\ \  i=\rm{Na^+},~\rm{K^+},~\rm{Cl^-}.$$
	
	Because of the direct connection, the communication ion flux $J_{l,k}^{i}$   between perivascular spaces only depends on transportation with fluid $C_{up,wind}^{i}U_{l,k},$
	
	$$J_{l,k}^{i}=C_{up,wind}^{i}U_{l,k};\ \  i=\rm{Na^+},~\rm{K^+},~\rm{Cl^-}.$$
	
	On the glial cell membranes, $J_{gl,k}^{c,i}$ is defined as
	
	\begin{equation}\label{Jtransmem_gl}
		J_{gl,k}^{c,i}=\dfrac{g_{gl}^{i}}{z^{i}e}(\phi_{gl}-\phi_{k}-E_{gl,k}^{i}),\ i=\rm{Na^+,K^+,Cl^-}, \ k=ex,pa,pv,pc,
	\end{equation}
	
	\noindent where $E_{gl,k}^{i}$ is the Nernst potential that describes the gradient of chemical potential in electrical units
	$$E_{gl,k}^{i}=\dfrac{k_BT}{ez^{i}}log\left(\dfrac{C_{k}^{i}}{C_{gl}^{i}}\right), k=ex,pa,pv,pc$$  and the conductance $g_{gl}^{i}$ for the $i$th ion species on the glial membrane is a fixed constant, independent of voltage and time. 
	
	On the axon's membrane, $J_{ax,ex}^{c,i}$ is definded as 
	
	\begin{equation}
		J_{ax,ex}^{c,i}=\dfrac{g_{ax}^{i}}{z^{i}e}(\phi_{ax}-\phi_{ex}-E_{ax}^{i}),\ i=\rm{Na^+,K^+,Cl^-},
	\end{equation}
	
	\noindent where the conductances of $\rm{Na^+}$ and $\rm{K^+}$ are modeled using  the Hodgkin-Huxley model \cite{1960Thresholds,2017Mathematics}
	
	$$g_{ax}^{Na}=\bar{g}^{Na}m^{3}h+g_{leak}^{Na},\ g_{ax}^{K}=\bar{g}^{K}n^{4}+g_{leak}^{K},\ g_{ax}^{Cl}=g_{leak}^{Cl}.$$
	
	\begin{equation}
		\begin{cases}
			\dfrac{dn}{dt}=\alpha_{n}(1-n)-\beta_{n}n,\\
			\dfrac{dm}{dt}=\alpha_{m}(1-m)-\beta_{m}m,\\
			\dfrac{dh}{dt}=\alpha_{h}(1-h)-\beta_{h}h,
		\end{cases}
	\end{equation}
	
	\noindent where $n$ is the open probability of $K^{+}$ channel, $m$ is the open probability of the $Na^{+}$ activation gate, and $h$ is the open probability of the $Na^{+}$ inactivation gate. $\alpha_i$ and $\beta_i$ for $i=n, m, h$ are active and inactive rates of different gate. 
	
	For the active ion pump source $J_{l,k}^{p,i}$, the only pump we consider is the Na/K active transporter. We are more than aware that other active transport systems can and likely do move ions and fluid in this system. They will be included as experimental information becomes available. In the case of the Na/K pump $J_{l,k}^{p,i},\ (l,k=ax,ex;gl,pa;gl,pv;gl,pc)$,  the strength of the pump depends on the concentration in the intracellular and extracellular spaces \cite{1960Thresholds,2000Isoform}, such that 
	
	\begin{equation}
		J_{l,k}^{p,Na}=\dfrac{3I_{l}}{e},\ J_{l,k}^{p,K}=-\dfrac{2I_{l}}{e},\ J_{l,k}^{p,Cl}=0,\ l=ax,gl,
	\end{equation}
	
	\noindent where 
	
	$$I_{l}=I_{l,1}\left(\dfrac{C_{l}^{Na}}{C_{l}^{Na}+K_{Na1}}\right)^{3}\left(\dfrac{C_{ex}^{K}}{C_{ex}^{K}+K_{K1}}\right)^{2}+I_{l,2}\left(\dfrac{C_{l}^{Na}}{C_{l}^{Na}+K_{Na2}}\right)^{3}\left(\dfrac{C_{ex}^{K}}{C_{ex}^{K}+K_{K2}}\right)^{2},$$
	
	\noindent $I_{l,1}$ and $I_{l,2}$ are related to $\alpha_{1}-$ and $\alpha_{2}-$ isoform of Na/K pump. 
	
	\paragraph{Ion Flux inside Compartment }	The definitions of ion flux in each domain are as follows, for $i=\rm{Na^+,~K^+,~Cl^-},$
	
	\begin{subequations}
		\begin{align}
			& \mathbf{j}_{l}^{i}=C_{l}^{i}\mathbf{u}_{l}-D_{l}^{i}\tau_{l}\left(\nabla C_{l}^{i}+\dfrac{z^{i}e}{k_{B}T}C_{l}^{i}\nabla\phi_{l}\right),\ l=gl,ex,pa,pv,pc, \\
			& j_{ax,z}^{i}=C_{ax}^{i}u_{ax}^{z}-D_{ax}^{i}\left(\dfrac{\partial C_{ax}^{i}}{\partial z}+\dfrac{z^{i}e}{k_{B}T}C_{ax}^{i}\dfrac{\partial\phi_{ax}}{\partial z}\right).\ \ \ \ \ \ \ \ \ \ \ \ \ \ \ \ \ \ \ \ \ \ \ \ \ \ \ \ \ \ \        
		\end{align}
	\end{subequations}
	\paragraph{Boundary Conditions} 	
	For the axon and glial compartment boundary condition, we use the membrane boundary conditions at location $\Gamma_2\cup\Gamma_6$. The homogeneous Neumann boundary condition on the $\Gamma_1$ and a non-flux boundary condition is used on the pia mater $\Gamma_7$ for the glial compartment.
	
	\begin{equation}
		\begin{cases}
			\nabla C_{gl}^{i}\cdot\hat{\mathbf{n}}_{r}=0, & \mbox{on} \ \Gamma_1,\\
			\nabla C_{ax}^{i}\cdot\hat{\mathbf{n}}_{z}=\lambda_{ax,left}(C_{ax}^{i}-C_{ax}^{i,re}),\ \nabla C_{gl}^{i}\cdot\hat{\mathbf{n}}_{z}=\lambda_{gl,left}(C_{gl}^{i}-C_{gl}^{i,re}), & \mbox{on} \ \Gamma_2,\\
			\nabla C_{ax}^{i}\cdot\hat{\mathbf{n}}_{z}=\lambda_{ax,right}(C_{ax}^{i}-C_{ax}^{i,re}),\ \nabla C_{gl}^{i}\cdot\hat{\mathbf{n}}_{z}=\lambda_{gl,right}(C_{gl}^{i}-C_{gl}^{i,re}), & \mbox{on} \ \Gamma_6,\\
			\mathbf{j}_{gl}^{i}\cdot\hat{\mathbf{n}}_{r}=0, & \mbox{on} \ \Gamma_7,\\
		\end{cases}
	\end{equation}
	where $\lambda_{ax(gl), left(right)}$ is the ion communication rate on the left (right) boundary of the axon (glial) compartment.

	We use the Dirichlet boundary conditions for the perivascular space A \& V boundary condition at location $\Gamma_1$. For the pvsC and the ECS boundary condition, we use the homogeneous Neumann boundary condition at location $\Gamma_1$. The membrane boundary conditions at locations $\Gamma_2\cup\Gamma_6$ are used for the perivascular space A \& V and the ECS. And the homogeneous Neumann boundary condition on $\Gamma_2\cup\Gamma_6$ is used for the pvsC. The flux across the pia mater is assumed continuous and Ohm's law \cite{2019ABidomain} and the additional pathway for diffusion, electric drift as well as convection for ions is used at location $\Gamma_7$ for the perivascular space A \& C \& V and the ECS.

	\begin{equation}
		\begin{cases}
			C_{pa}^{i}=C_{pa}^{i,re},\ C_{pv}^{i}=C_{pv}^{i,re},\ \nabla C_{pc}^{i}\cdot\hat{\mathbf{n}}_{r}=0,\ \nabla C_{ex}^{i}\cdot\hat{\mathbf{n}}_{r}=0, & \mbox{on} \ \Gamma_1,\\
			\nabla C_{pa}^{i}\cdot\hat{\mathbf{n}}_{z}=\lambda_{pa,left}(C_{pa}^{i}-C_{pa}^{i,re}),\ \nabla C_{pv}^{i}\cdot\hat{\mathbf{n}}_{z}=\lambda_{pv,left}(C_{pv}^{i}-C_{pc}^{i,re}), & \mbox{on} \ \Gamma_2,\\
			\nabla C_{pc}^{i}\cdot\hat{\mathbf{n}}_{z}=\lambda_{pc,left}(C_{pc}^{i}-C_{pc}^{i,re}),\ \nabla C_{ex}^{i}\cdot\hat{\mathbf{n}}_{z}=\lambda_{ex,left}(C_{ex}^{i}-C_{ex}^{i,re}), & \mbox{on} \ \Gamma_2,\\
			\nabla C_{pa}^{i}\cdot\hat{\mathbf{n}}_{z}=\lambda_{pa,right}(C_{pa}^{i}-C_{pa}^{i,re}),\ \nabla C_{pv}^{i}\cdot\hat{\mathbf{n}}_{z}=\lambda_{pv,right}(C_{pv}^{i}-C_{pc}^{i,re}), & \mbox{on} \ \Gamma_6,\\
			\nabla C_{pc}^{i}\cdot\hat{\mathbf{n}}_{z}=\lambda_{pc,right}(C_{pc}^{i}-C_{pc}^{i,re}),\ \nabla C_{ex}^{i}\cdot\hat{\mathbf{n}}_{z}=\lambda_{ex,right}(C_{ex}^{i}-C_{ex}^{i,re}), & \mbox{on} \ \Gamma_6,\\
			\mathbf{j}_{pa}^{i}\cdot\hat{\mathbf{n}}_{r}=\dfrac{G_{pia}^{i}}{z^{i}e}\left(\phi_{pa}-\phi_{csf}-\dfrac{k_BT}{ez^{i}}log\left(\dfrac{C_{csf}^{i}}{C_{pa}^{i}}\right)\right)+C_{pa}^{i}u_{pa,csf}, & \mbox{on} \ \Gamma_7,\\
			\mathbf{j}_{pv}^{i}\cdot\hat{\mathbf{n}}_{r}=\dfrac{G_{pia}^{i}}{z^{i}e}\left(\phi_{pv}-\phi_{csf}-\dfrac{k_BT}{ez^{i}}log\left(\dfrac{C_{csf}^{i}}{C_{pv}^{i}}\right)\right)+C_{pv}^{i}u_{pv,csf}, & \mbox{on} \ \Gamma_7,\\
			\mathbf{j}_{pc}^{i}\cdot\hat{\mathbf{n}}_{r}=\dfrac{G_{pia}^{i}}{z^{i}e}\left(\phi_{pc}-\phi_{csf}-\dfrac{k_BT}{ez^{i}}log\left(\dfrac{C_{csf}^{i}}{C_{pc}^{i}}\right)\right), & \mbox{on} \ \Gamma_7,\\
			\mathbf{j}_{ex}^{i}\cdot\hat{\mathbf{n}}_{r}=\dfrac{G_{pia}^{i}}{z^{i}e}\left(\phi_{ex}-\phi_{csf}-\dfrac{k_BT}{ez^{i}}log\left(\dfrac{C_{csf}^{i}}{C_{ex}^{i}}\right)\right), & \mbox{on} \ \Gamma_7.\\
		\end{cases}
	\end{equation}
	
	For the cerebrospinal fluid boundary condition, similar boundary conditions are imposed except on $\Gamma_5$, which a non-permeable boundary condition is used.
	
	\begin{equation}
		\begin{cases}
			C_{csf}^{i}=C_{csf}^{i,re}, & \mbox{on} \ \Gamma_3,\\
			\nabla C_{csf}^{i}\cdot\hat{\mathbf{n}}_{r}=0, & \mbox{on} \ \Gamma_4,\\
			\mathbf{j}_{csf}^{i}\cdot\hat{\mathbf{n}}_{z}=0, & \mbox{on} \ \Gamma_5,\\
			\mathbf{j}_{csf}^{i}\cdot\hat{\mathbf{n}}_{r}= \sum\limits_{l=pa,pv,pc,ex} \mathbf{j}_{l}^{i}\cdot\hat{\mathbf{n}}_{r}& \mbox{on} \ \Gamma_7.
		\end{cases}
	\end{equation}
	
	\subsection{Electric Potential}
	
	By multiplying equations \ref{eq:Iontransport} with $z_{i}e$ respectively, summing up, and using charge neutrality equation \ref{Chargeneutrality} and ion flux equation, we have following system for the electric potential in $ax,gl,ex,pa,pv,pc$
	
	\begin{subequations}
		\begin{align}
			& \sum_{i}z^{i}e\left(\mathcal{M}_{ax,ex}J_{ax,ex}^{i}+\dfrac{\partial}{\partial z}(\eta_{ax}j_{ax,z}^{i})\right)=0,\\
			\begin{split}
				& \sum_{k=ex,pa,pv,pc}\left(\sum_{i}z^{i}e\left(\mathcal{M}_{gl,k}J_{gl,k}^{i}\right)\right)+\sum_{i}z^{i}e\left(\nabla\cdot(\eta_{gl}\mathbf{j}_{gl}^{i})\right)=0,   \\
			\end{split}\\
			\begin{split}
				& \sum_{i}z^{i}e\left(\mathcal{M}_{pa,ex}J_{pa,ex}^{i}-\mathcal{M}_{gl,pa}J_{gl,pa}^{i}+\mathcal{M}_{pa,pc}J_{pa,pc}^{i}+\nabla\cdot(\eta_{pa}\mathbf{j}_{pa}^{i})\right)=0,\\
			\end{split}\\
			\begin{split}
				& \sum_{i}z^{i}e\left(\mathcal{M}_{pv,ex}J_{pv,ex}^{i}-\mathcal{M}_{gl,pv}J_{gl,pv}^{i}-\mathcal{M}_{pc,pv}J_{pc,pv}^{i}+\nabla\cdot(\eta_{pv}\mathbf{j}_{pv}^{i})\right)=0,\\
			\end{split}\\
			\begin{split}
				& \sum_{i}z^{i}e\left(\mathcal{M}_{pc,ex}J_{pc,ex}^{i}-\mathcal{M}_{gl,pc}J_{gl,pc}^{i}-\mathcal{M}_{pa,pc}J_{pa,pc}^{i}+\mathcal{M}_{pc,pv}J_{pc,pv}^{i}+\nabla\cdot(\eta_{pc}\mathbf{j}_{pc}^{i})\right)=0,\
			\end{split}\\
			\begin{split}
				& \sum_{k=ax,gl,pa,pv,pc}\left(-\sum_{i}z^{i}e\left(\mathcal{M}_{k,ex}J_{k,ex}^{i}\right)\right)+\sum_{i}z^{i}e\left(\nabla\cdot(\eta_{ex}\mathbf{j}_{ex}^{i})\right)=0,\\
			\end{split}
		\end{align}
	\end{subequations}
	
	\noindent which describe the spatial distributions of electric potentials in six compartments.
	
	In the subarachnoid space $\Omega_{SAS}$,  the governing equation for cerebrospinal fluid electric potential reduces to
	
	\begin{equation}
		\sum_{i}z^{i}e\left(\nabla\cdot\mathbf{j}_{csf}^{i}\right)=0.
	\end{equation}
	
	The boundary conditions for electric fields $\phi_{ax},\phi_{gl},\phi_{pa},\phi_{pv},\phi_{pc},\phi_{ex},\phi_{csf}$ are given below.
	
	\begin{equation}
		\begin{cases}
			\nabla\phi_{l}\cdot\hat{\mathbf{n}}_{r}=0,\ l=gl,pa,pv,pc,ex, & \mbox{on} \ \Gamma_1,\\
			\nabla\phi_{l}\cdot\hat{\mathbf{n}}_{z}=0,\ l=ax,gl,pa,pv,pc,ex, & \mbox{on} \ \Gamma_2\cup\Gamma_6,\\
			\nabla\phi_{csf}\cdot\hat{\mathbf{n}}_{r}=0, & \mbox{on} \ \Gamma_3\cup\Gamma_5,\\
			\nabla\phi_{csf}\cdot\hat{\mathbf{n}}_{z}=0, & \mbox{on} \ \Gamma_4,\\
			\nabla\phi_{gl}\cdot\hat{\mathbf{n}}_{r}=0, & \mbox{on} \ \Gamma_7,\\
			\sum_{i}z^{i}e\mathbf{j}_{l}^{i}\cdot\hat{\mathbf{n}}_{r}=\sum_{i}G_{pia}^{i}\left(\phi_{l}-\phi_{csf}-\dfrac{k_BT}{ez^{i}}log\left(\dfrac{C_{csf}^{i}}{C_{l}^{i}}\right)\right),\ l=pa,pv,pc,ex, & \mbox{on} \ \Gamma_7,\\
			\sum_{i}z^{i}e\mathbf{j}_{csf}^{i}\cdot\hat{\mathbf{n}}_{r}= \sum\limits_{l=pa,pv,pc,ex}\sum_{i}z^{i}e\mathbf{j}_{l}^{i}\cdot\hat{\mathbf{n}}_{r}& \mbox{on} \ \Gamma_7.\\
		\end{cases}
	\end{equation}

	\section{Results }
	
	In this section, we present the simulation results obtained using the proposed model to understand microcirculation in the optic nerve. Specifically, we focus on how glial cells and perivascular spaces facilitate potassium and metabolic waste clearance during and after a stimulus.
	
	The model is solved using the Finite Volume Method, with a mesh size of $h=1/20$ and a temporal size of $t=1/10$ in dimensionless units. The code was developed and executed in the Matlab environment. Initially, the resting state of the system is determined through iteration and by setting a fixed volume fraction \cite{1966Physiological}:
	
	\[\eta_{ax}^{re}=0.4, \eta_{gl}^{re}=0.4, \eta_{ex}^{re}=0.1, \eta_{pa}^{re}=0.024, \eta_{pv}^{re}=0.0639, \eta_{pc}^{re}=0.0121.\]
	Then, these resting-state values are used as initial conditions for the stimulus state. 
	
\subsection{Micro-circulation in the Resting State}



Under a pressure gradient of $0.0083~\rm{mmHg/mm}$, cerebrospinal fluid (CSF) flows into the subarachnoid space (SAS) from the intracranial region, with an average velocity of approximately $250 ~\rm{\mu m/s}$ in the z-direction, consistent with previous findings \cite{2021Lymphatics}. CSF then leaks into perivascular space A. The flow direction in perivascular space A (or V) matches the blood flow in the central retinal artery (or vein). The fluid flow in perivascular space C adjusts dynamically through exchanges with other compartments, primarily influenced by perivascular space V due to a larger pressure gradient at boundary $\Gamma_1$, with fluid exiting the optic nerve at boundary $\Gamma_2$ under membrane boundary conditions.

In perivascular space A, the spatially averaged fluid velocity under a $0.007~\rm{mmHg/mm}$ pressure gradient in the z-direction is $5~\rm{ \mu m/s}$ from the intracanalicular space to the intraorbital region. In perivascular space V, the spatially averaged fluid velocity in the z-direction under a $-0.012 ~\rm{mmHg/mm}$ pressure gradient is $4~\mu \rm{m/s}$ from the intraorbital region to the intracanalicular space. These results align with the observations in \cite{2022Arterial, boster2023artificial}.

Simultaneously, fluid flows from pvsA into the ECS and the glial compartment through gaps and Aquaporin channels located near or on the glial cell feet \cite{2017Aquaporin}, and then into perivascular space V. Fluid also moves from the ECS into the glial compartment, in agreement with the glymphatic system’s role \cite{2015TheGlymphatic, 2017Evidence, 2020Glymphatic}.

\subsection{Micro-circulation in the Stimulated State}

\begin{figure}
    \centering
    \includegraphics[width=0.6\linewidth]{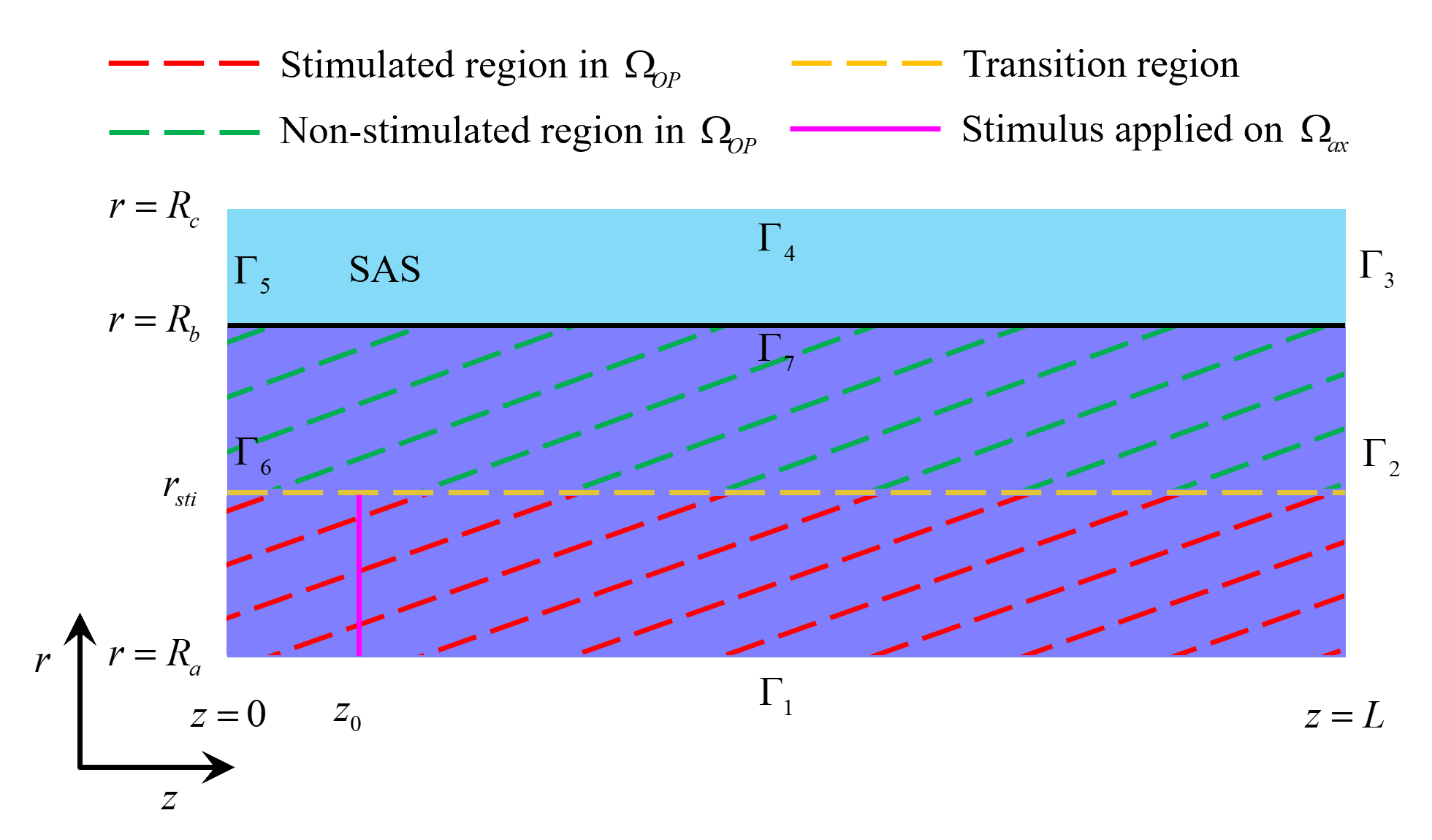}
    \caption{\label{fig:Sti_regin} Stimulated and non-stimulated regions in the optic nerve ($\Omega_{OP}$). The stimulus is applied to the axon membrane within the region $R_{a}<r<r_{sti}$ at a specified location $z=z_{0}$.}
\end{figure}

This section examines the glymphatic system's role (including glial cells and perivascular spaces) in metabolic waste clearance during stimulation. As depicted in Fig. \ref{fig:Sti_regin}, the stimulus is applied to the axon membrane within the region $R_{a}<r<r_{sti}<R_b$ at a specified location $z=z_{0}$, with $\theta\in[0,2\pi]$. We differentiate between the stimulated and non-stimulated regions in the optic nerve $\Omega_{OP}$, where electrical signals propagate in the z-direction within the axon compartment. The stimulus frequency is $50\ \rm{Hz}$ ($T=0.02\ \rm{s}$) with a duration of $0.2\ \rm{s}$. Each stimulus has a current strength of $I_{sti}=3\times10^{-3}\ \rm{A/m^{2}}$ and lasts for $3\ \rm{ms}$.

\subsubsection{Ionic Circulation and Potassium Clearance}

\begin{figure}
    \centering
    \includegraphics[width=1\linewidth]{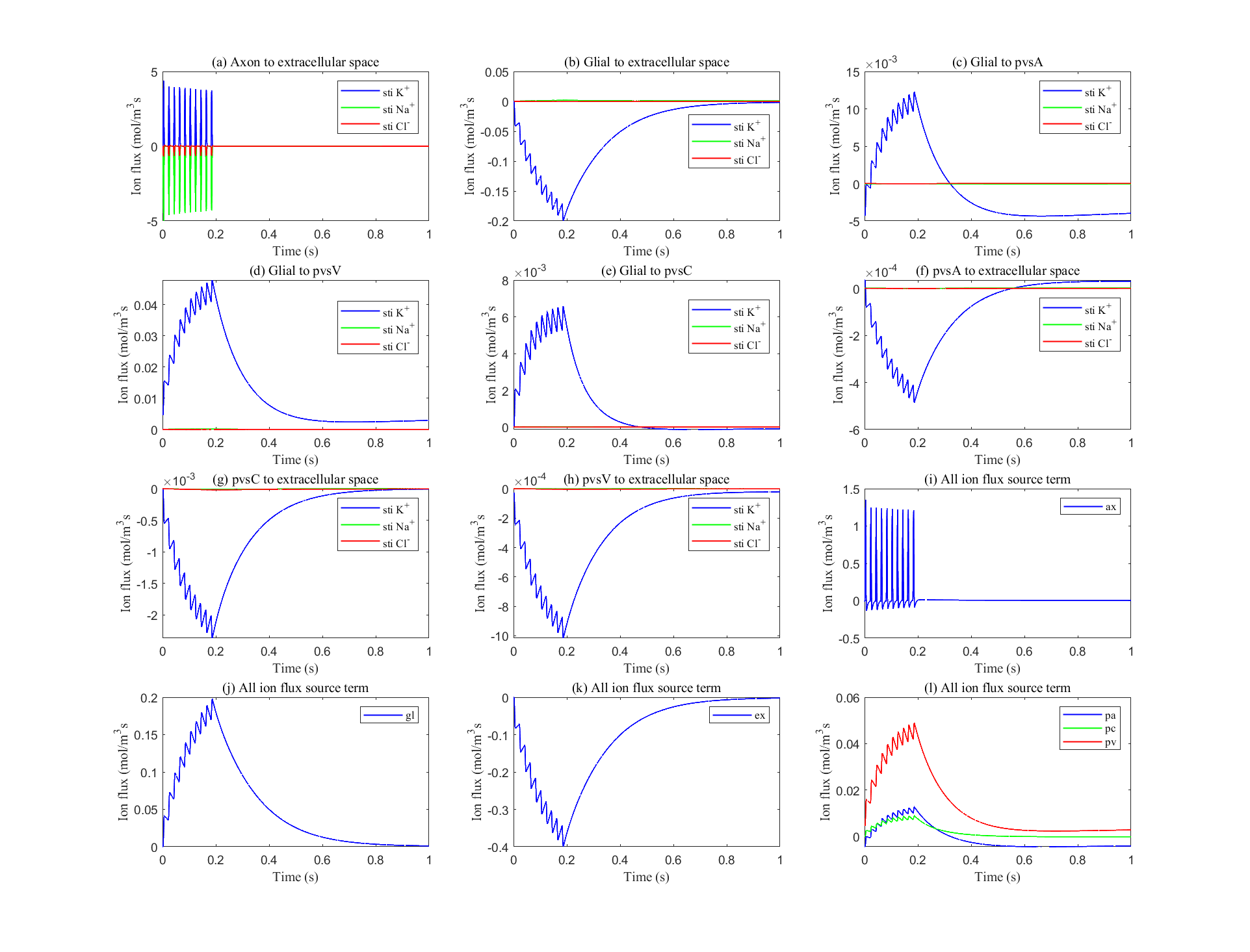}
    \caption{\label{fig:AllFluxTerm} Numerical results. Panels a-h: Average transmembrane ion flux in the stimulated region. Panels i-l: Average sum of all ion transmembrane source terms for each compartment, with positive values indicating increased osmotic concentration and negative values indicating decreased osmotic concentration.}
\end{figure}

\begin{figure}
    \centering
    \includegraphics[width=0.8\linewidth]{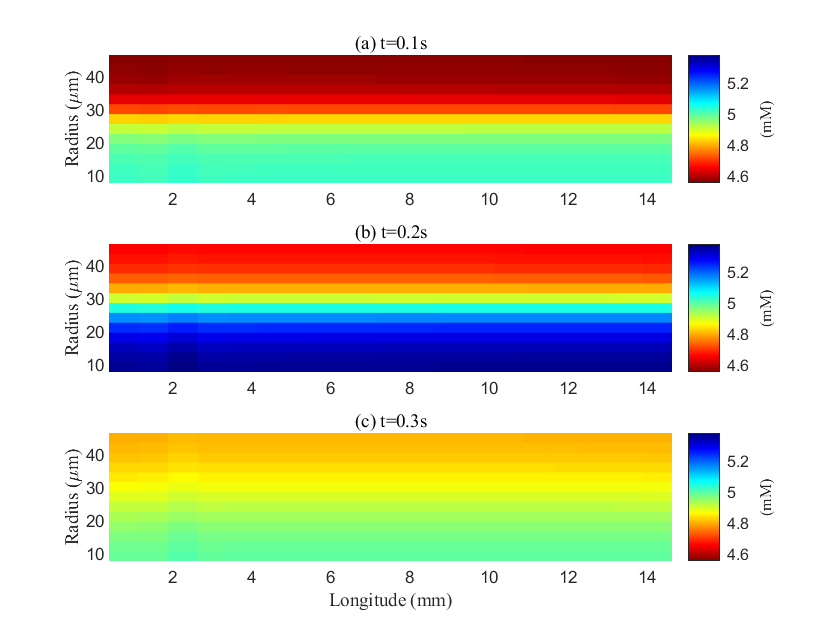}
    \caption{\label{fig:KConAllsti} Spatial distribution of potassium concentration during and after a train of stimuli in the ECS.}
\end{figure}

This section explores ionic circulation between stimulated and non-stimulated regions. During stimulation, as shown in Fig. \ref{fig:AllFluxTerm}a, $\rm{Na^+}$ ions enter the axon while $\rm{K^+}$ ions leak from the axon into the ECS, leading to potassium accumulation in the stimulated region. Snapshots of potassium concentration in the ECS, shown in Fig. \ref{fig:KConAllsti}, indicate that potassium concentration initially rises in the stimulated region, reaching up to $5.3 \ \rm{mM}$. Subsequently, through the buffering function of the glymphatic system, the accumulated potassium is transported from the stimulated to the non-stimulated region, lowering the overall concentration in the optic nerve to $4.9 \ \rm{mM}$. 

\begin{figure}
    \centering
    \includegraphics[width=0.8\linewidth]{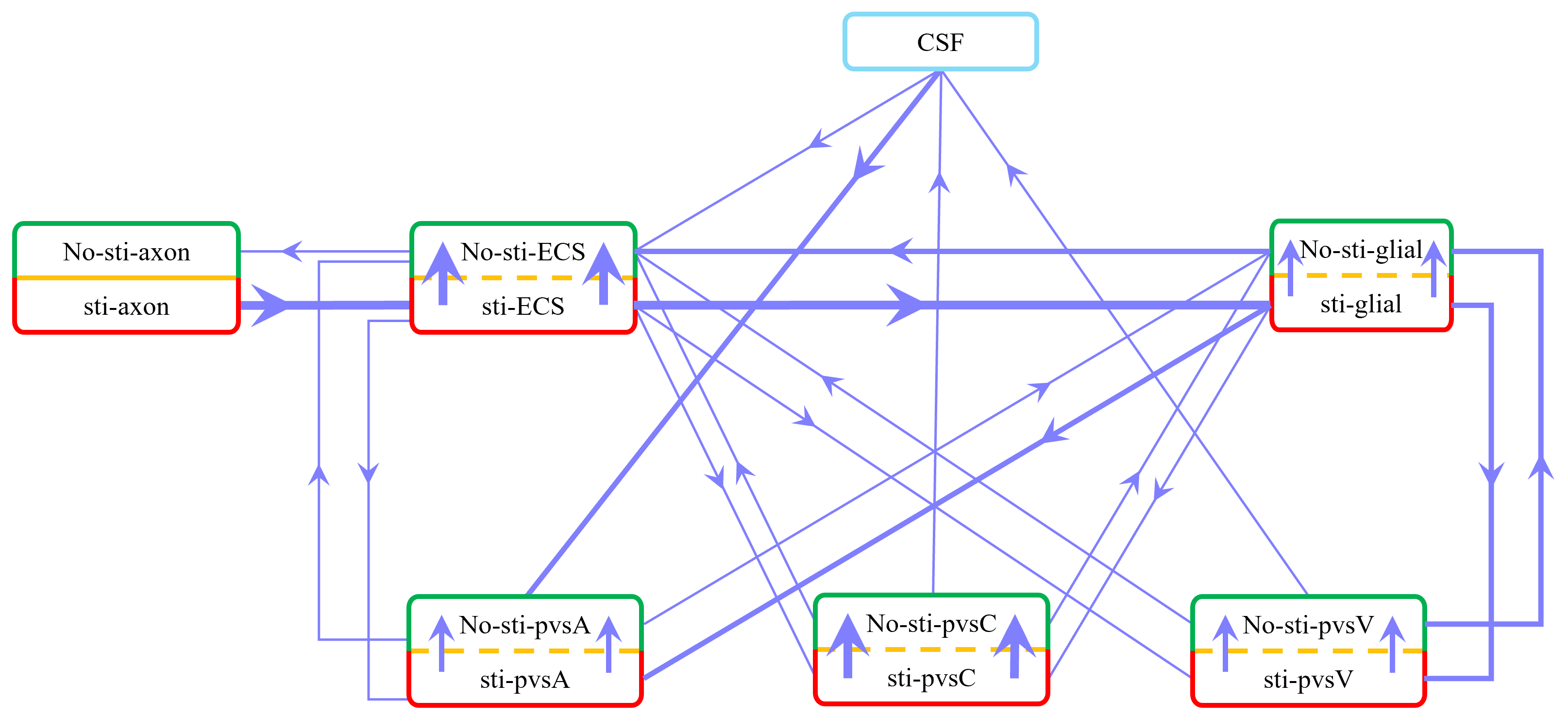}
    \caption{\label{fig:TransIonFlux} Schematic of potassium flux between the stimulated (lower) and non-stimulated (upper) regions, as well as transmembrane flux between different compartments during stimulation. Red boxes represent stimulated regions, and green boxes represent non-stimulated regions. The thickest lines indicate fluxes around $\rm{10^{-1} ~mol/(m^{3}s)}$, moderately thick lines represent fluxes around $\rm{10^{-2} ~mol/(m^{3}s)}$, and the thinnest lines indicate fluxes less than $\rm{10^{-3} ~mol/(m^{3}s)}$.}
\end{figure}

Fig. \ref{fig:TransIonFlux} illustrates three mechanisms for buffering the accumulated potassium: the glial compartment, the perivascular space, and diffusion within the ECS.

\begin{figure}
    \centering
    \includegraphics[width=1\linewidth]{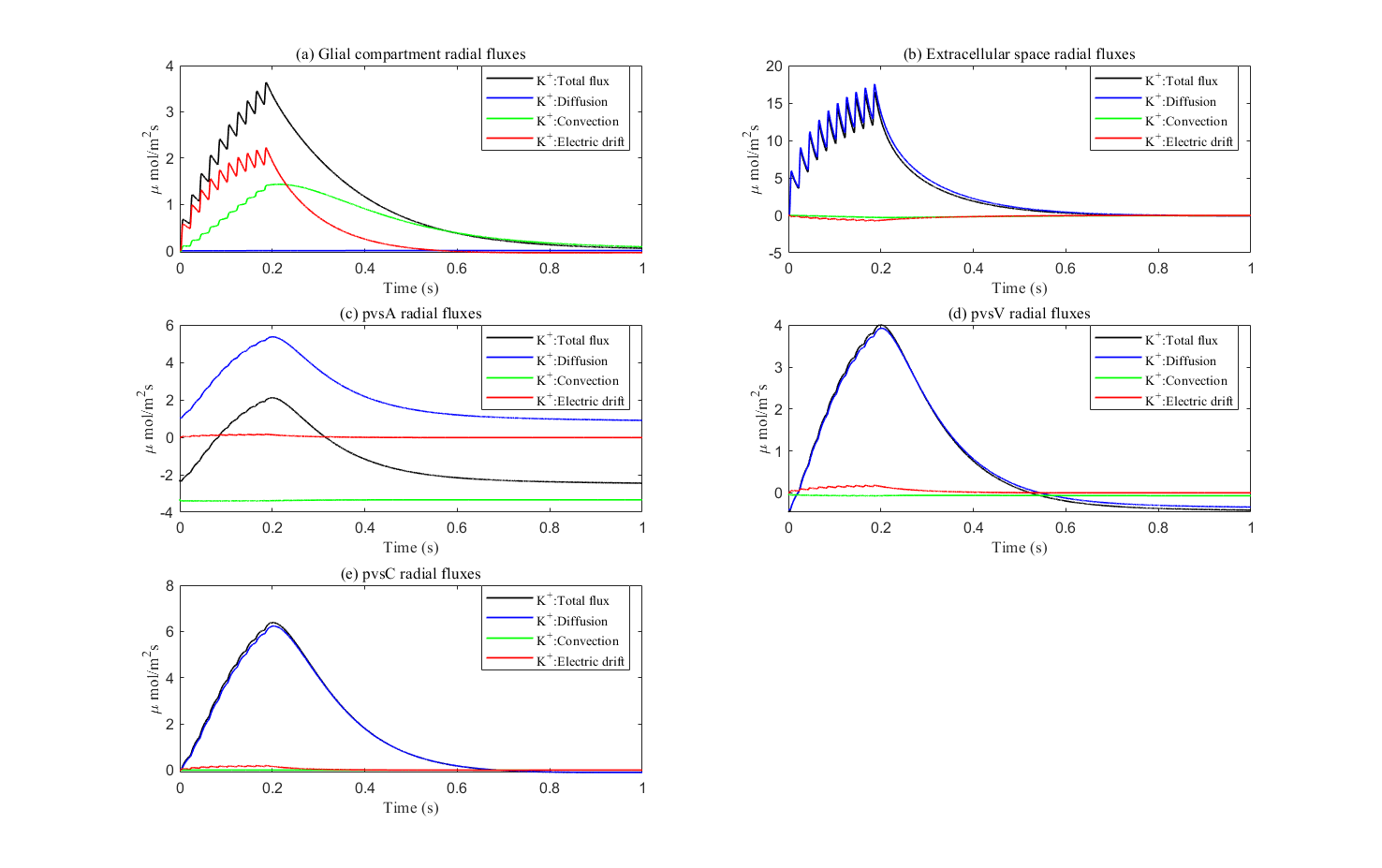}
    \caption{\label{fig:Diffusion} Average radial direction potassium flux components within each compartment.}
\end{figure}

\begin{figure}
    \centering
    \includegraphics[width=1\linewidth]{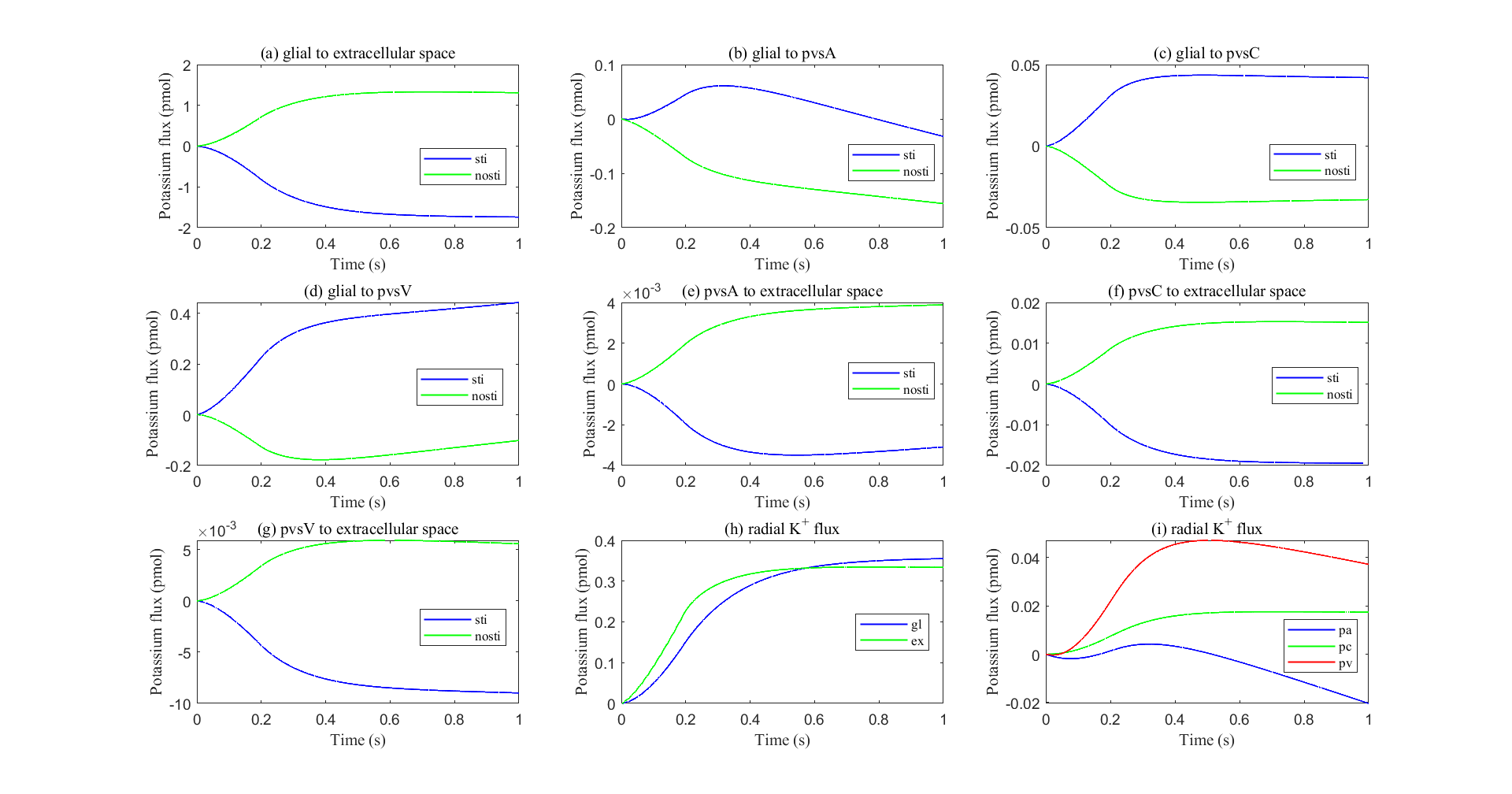}
    \caption{\label{fig:TotalIonFluxCum} Panels a-g: Cumulative transmembrane potassium fluxes in the stimulated and non-stimulated regions after neuronal firing stops. Panels h-i: Radial cumulative potassium fluxes within compartments after neuronal firing stops.}
\end{figure}

\begin{itemize}
    \item \textbf{Glial Compartment:} The glial compartment is the primary pathway for potassium buffering, protecting axons. Potassium accumulation in the stimulated region generates a Nernst potential, $E_{gl,ex}^{k}$, across the glial membrane, driving potassium flow from the extracellular space into the glial compartment, as shown in Fig. \ref{fig:AllFluxTerm}b. This rapid influx raises the local electric potential, facilitating potassium redistribution from the stimulated to the non-stimulated region. Due to connexins, the glial compartment behaves as an electrical syncytium, causing the membrane potential in the non-stimulated region to also become more positive. However, the glial potassium Nernst potential in the non-stimulated region remains near its resting state, resulting in an outward potassium flux from the glial compartment. Fig. \ref{fig:Diffusion}a shows that during the stimulus, electric drift flux primarily buffers potassium, while post-stimulus, convection flux driven by water circulation becomes dominant.

    \item \textbf{Perivascular Space:} The perivascular space serves as a secondary pathway for potassium buffering. Potassium leaks from the ECS into the perivascular space through gaps between astrocytes' endfeet (see Fig. \ref{fig:AllFluxTerm}f-h), and some potassium from the glial compartment also enters via channels on astrocytes' endfeet (see Fig. \ref{fig:AllFluxTerm}g-e). Potassium is transported from the stimulated region to the non-stimulated region through convection and diffusion. In the non-stimulated region, potassium leaks back from the perivascular space into the ECS. Fig. \ref{fig:Diffusion}c-e shows that diffusion flux plays a key role in potassium buffering in the perivascular spaces. Convective flow driven by pressure differences transports potassium from the SAS to the inner optic nerve. Since potassium concentrations in perivascular spaces A, V, and C are much lower than in the glial compartment, the contribution of electric drift to potassium buffering is negligible. During stimulation, perivascular spaces A, V, and C serve as additional pathways for potassium transport from the stimulus region to the non-stimulus region. After stimulation, perivascular spaces V and C continue buffering potassium, while in perivascular space A, convection flux suppresses diffusion, leading to potassium flux directed from the pia mater to the optic nerve center, depending on the stimulation location (see Fig... in the attachment).

    \item \textbf{Diffusion in the ECS:} As shown in Fig. \ref{fig:Diffusion}b, the connectivity of the ECS facilitates direct diffusion of accumulated potassium from the stimulated to the non-stimulated region. Fig. \ref{fig:TotalIonFluxCum}h illustrates that the total accumulated $K^+$ buffering through ECS via diffusion is comparable with the glial compartment. 
\end{itemize}

A detailed illustration of potassium microcirculation between the different compartments is available in Appendix Fig. \ref{fig:IonTransition}. And the sodium flux inside each domain can be found in Appendix Fig. \ref{fig:NaDiffusion}. 

Figure \ref{fig:TotalIonFluxCum} presents the cumulative potassium flux (total potassium flux integrated over time). The results indicate that the cumulative potassium flux through the glial membrane is twice as large as the radial cumulative potassium flux inside the ECS, two orders of magnitude larger than in pvsC, and three orders of magnitude larger than in pvsA and pvsV within the stimulated ECS. Concurrently, the cumulative potassium flux through the glial membrane into pvsV is 1.5 times the radial cumulative potassium flux inside the glial membrane and an order of magnitude larger than in pvsA and pvsC within the stimulated glial compartment during a train of stimuli.

In summary, these numerical results confirm that the glial compartment is the most critical and rapid pathway for potassium transport, while the perivascular spaces provide secondary pathways for potassium removal from the stimulated region when neurons fire or stop firing. Over time, potassium circulation and concentrations in each compartment return to the resting state.

	\subsubsection{Fluid Circulation}
While this paper primarily focuses on ionic microcirculation, we provide a brief summary of fluid circulation, with further details to be explored in a separate paper. As illustrated in Appendix Fig. \ref{fig:TransFluidFlux}, fluid circulation during and after stimulation is driven by two primary forces: hydrostatic pressure differences and osmotic pressure differences. Hydrostatic pressure gradients within the subarachnoid space (SAS) and perivascular spaces direct the movement of cerebrospinal fluid (CSF) and interstitial fluid through the optic nerve compartments. Simultaneously, osmotic pressure, influenced by ionic microcirculation, drives fluid exchange between the extracellular space (ECS), glial compartments, and perivascular spaces. These mechanisms work together to facilitate fluid redistribution, ensuring efficient clearance of metabolic waste and maintaining homeostasis between the stimulated and non-stimulated regions. The dynamic interaction between these forces allows fluid circulation to adapt in response to neuronal activity, supporting the glymphatic system's overall function in the optic nerve.

\subsection{Effects of Glial Compartments on Potassium Clearance}

In this section, we investigate the role of the glial compartment in extracellular potassium buffering by varying the transmembrane ionic conductance ($g_{gl}^i$), hydraulic permeability, and modifying connexin connectivity, which reduces diffusion coefficients and permeability within the glial compartment.

\subsubsection{Membrane ionic conductivity}
\begin{figure}
		\centering
		\includegraphics[width=\linewidth]{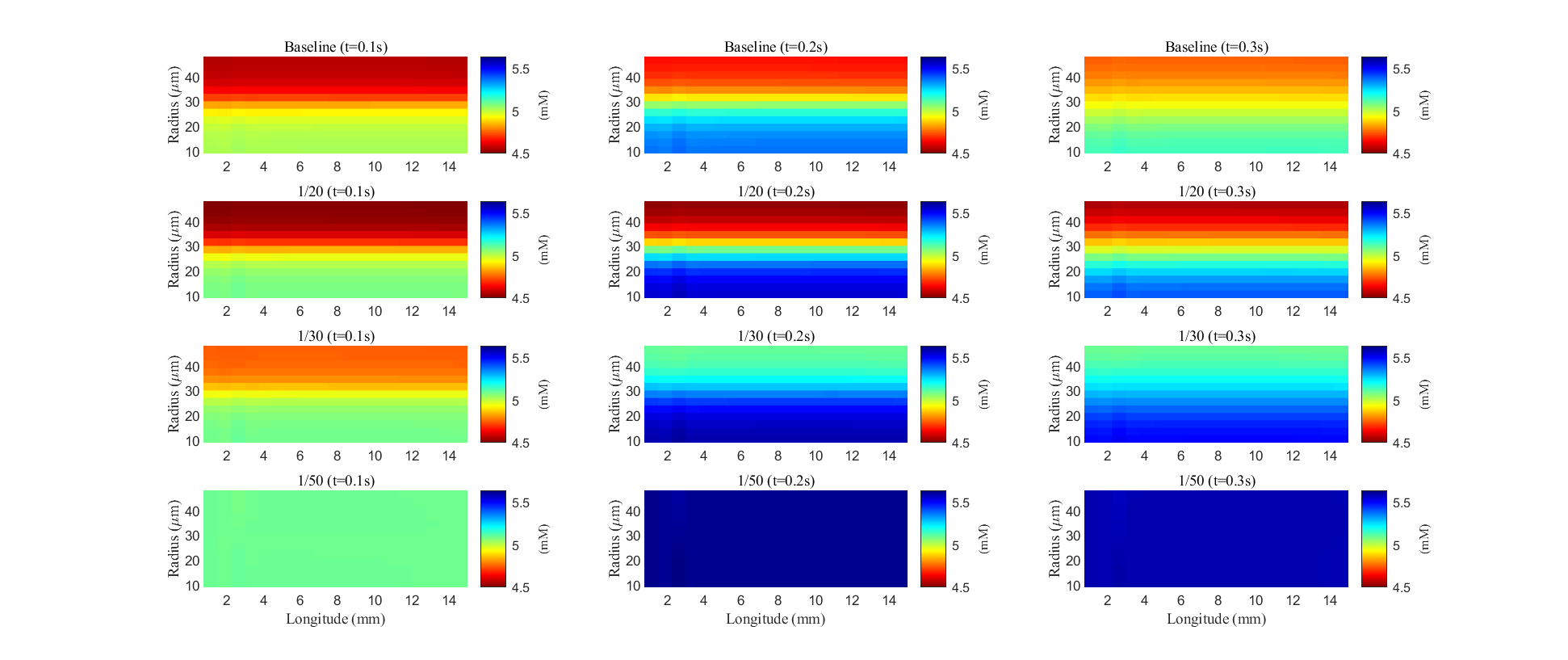}
		\caption{\label{fig:ConSpaceCase123} Spatial distribution of potassium concentration during and after a train of stimuli in the ECS. Different rows are results with different membrane conductance; Different columns are results at different time slots.}
    \end{figure}

Research suggests that in neurodegenerative diseases, oxidative stress and inflammation can impair ion channels on astrocytes, reducing their conductivity \cite{olufunmilayo_oxidative_2023,orfali_oxidative_2024}. To simulate these effects, we reduced the membrane conductivity to $\frac{1}{20}, \frac{1}{30}, \frac{1}{50}$ of the original values. Figure \ref{fig:ConSpaceCase123} shows the spatial distribution of extracellular potassium at various time points for different membrane conductance levels. As previously discussed, potassium clearance occurs via three main pathways: diffusion in the ECS, buffering by the glial compartment, and transport through the perivascular spaces.

\begin{figure}
    	\centering
    	\includegraphics[width=1\linewidth]{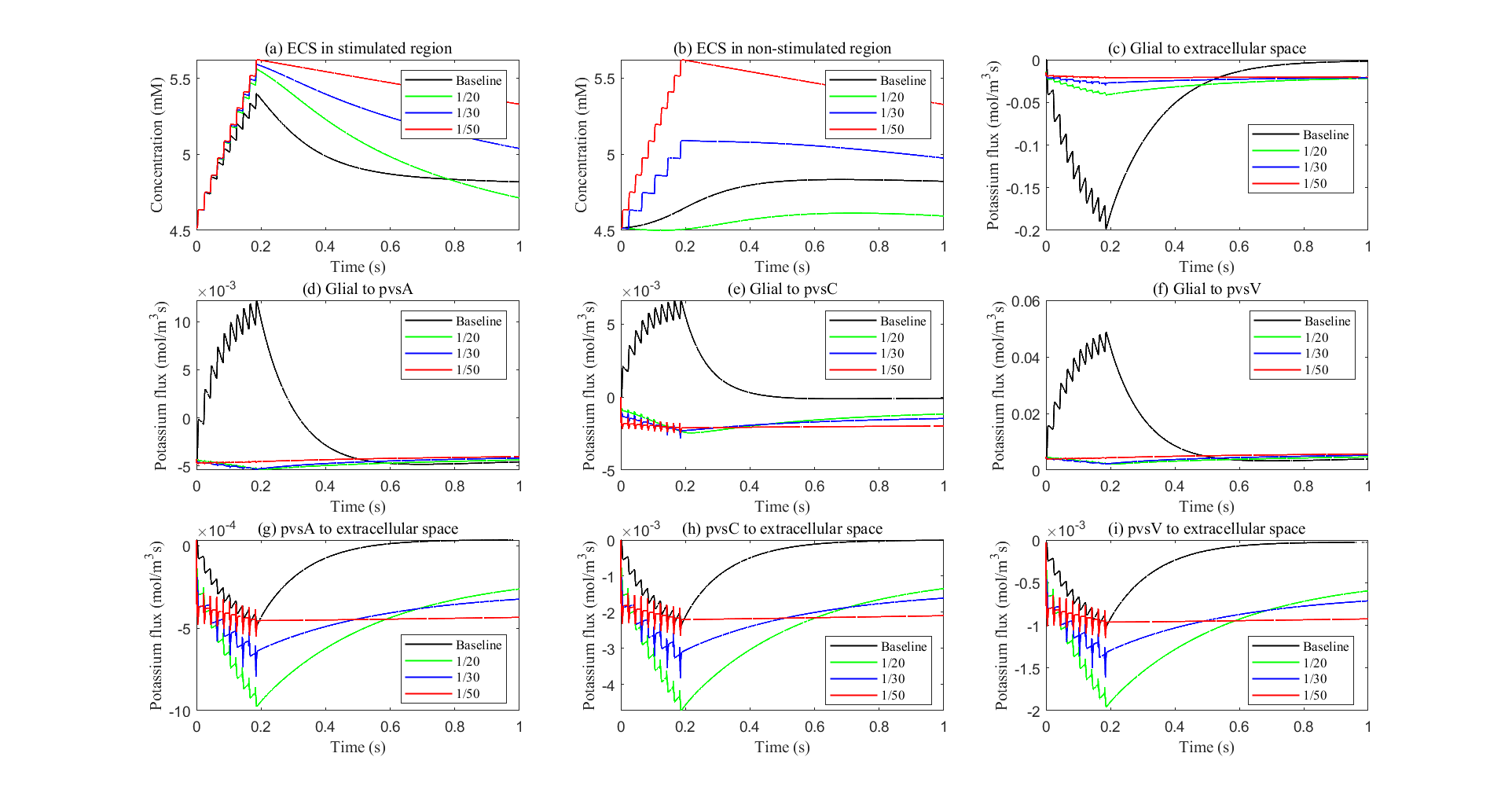}
    	\caption{\label{fig:KConTrans4} Potasium concentration and flux with different glial membrane conductance. a-b: The potassium concentration in the stimulated and non-stimulated regions. c-i: The transmembrane potassium flux in stimulated region.}
    \end{figure}

\begin{figure}
		\centering
		\includegraphics[width=0.8\linewidth]{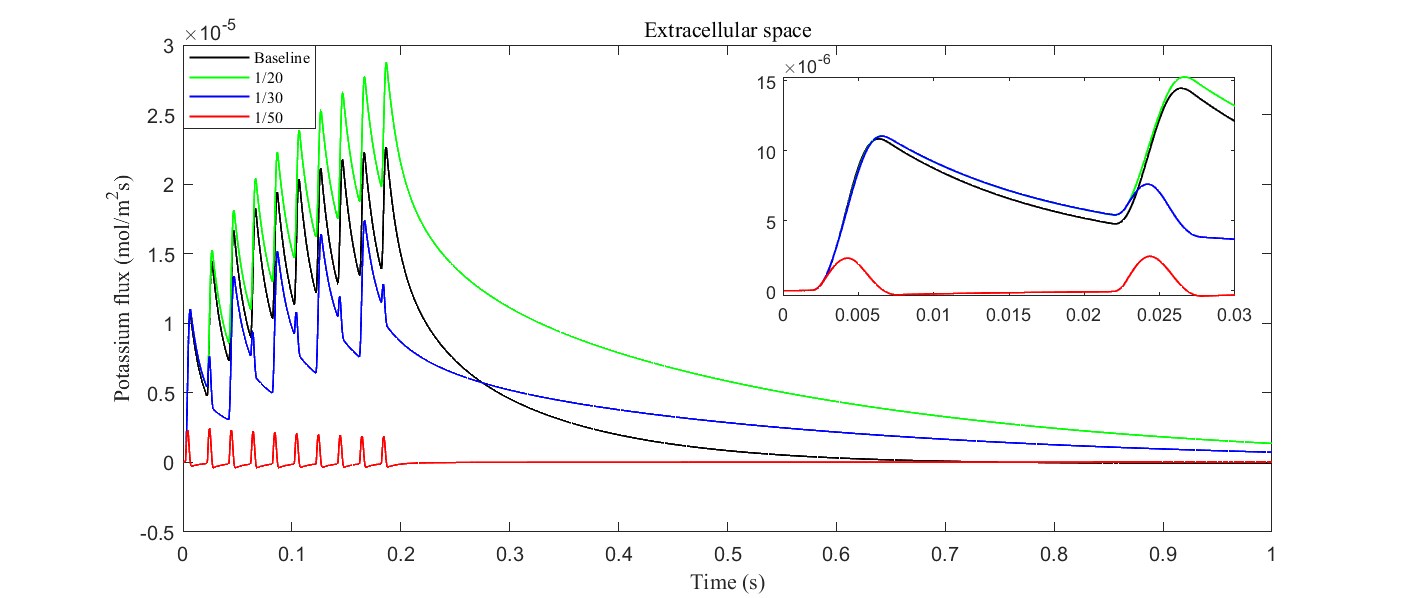}
	\caption{\label{fig:EcsInnerRIonFlux4}   Average radial potassium flux within the ECS with different glial membrane conductance.}
    \end{figure}

As glial membrane conductance decreases, the transmembrane flux through the glial compartment is reduced (see Fig. \ref{fig:KConTrans4}c-f), resulting in less efficient potassium clearance from the ECS (see Fig. \ref{fig:KConTrans4}a). When the conductance is reduced to $\frac{1}{20}$ of its original value, the perivascular spaces compensate by increasing their transmembrane flux, becoming the primary mechanism for potassium clearance (see Fig. \ref{fig:KConTrans4}g-i). This reduces the rate of potassium accumulation in the non-stimulated region (see Fig. \ref{fig:KConTrans4}b). However, the difference in potassium concentration between the stimulated and non-stimulated regions increases, leading to enhanced diffusion in the ECS (see Fig. \ref{fig:EcsInnerRIonFlux4}).

\begin{figure}
    	\centering
    	\includegraphics[width=1\linewidth]{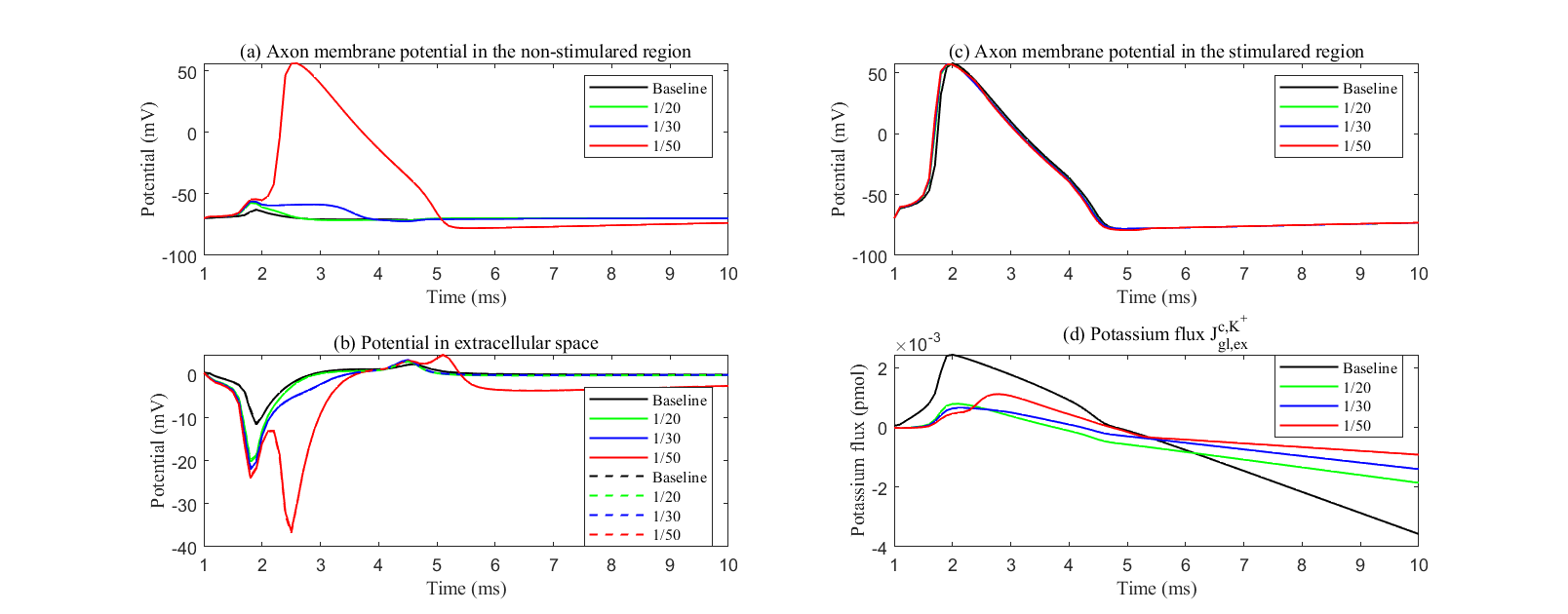}
    	\caption{\label{fig:PotentialFlux4} a: Axon membrane potential in the non-stimulated region. b: The electric potential in the ECS in the stimulated region (solid line) and non-stimulated region (dash line). d: The cumulative potassium flux through the passive ion channel on the glial membrane from the glial compartment to the ECS. Different lines mean different glial membrane conductance.}
    \end{figure}

The role of the glial compartment is not limited to potassium buffering. Figure \ref{fig:PotentialFlux4} shows the results during the first stimulus. At the onset of stimulation, the field potential in the extracellular region decreases (see Fig. \ref{fig:PotentialFlux4}a) as $\rm{Na^+}$ channels activate, allowing more $\rm{Na^+}$ to enter the axon. This raises the axonal potential and causes potassium to leak out of the glial compartment because the transmembrane potential ($\phi_{gl} - \phi_{ex}$) is higher than the Nernst potential ($E_{gl,k}^{ex}$) (see Fig. \ref{fig:PotentialFlux4}d). This process prevents the extracellular potential from dropping too low.

\begin{figure}
    	\centering
    	\includegraphics[width=1\linewidth]{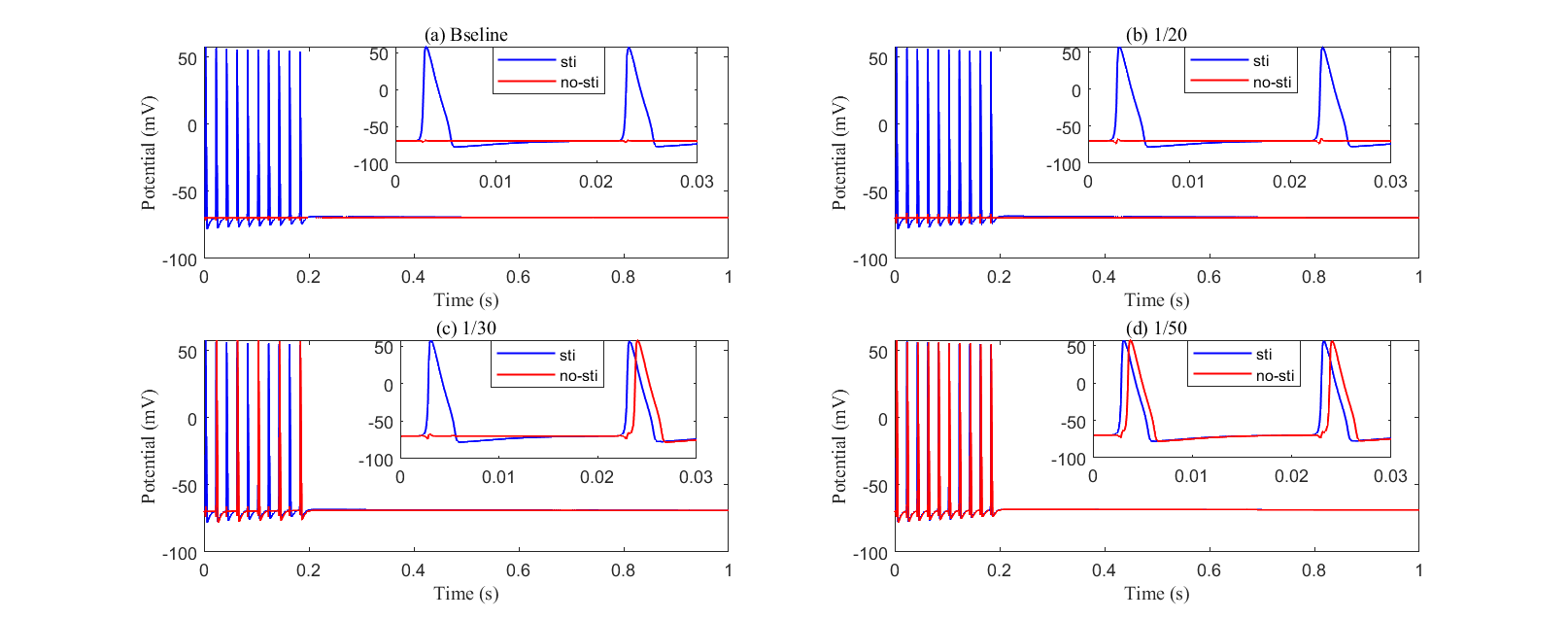}
    	\caption{\label{fig:MembranePtential4} Recording axon membrane potential in the stimulated and non-stimulated region with different glial membrane conductance.}
    \end{figure}

During the later stages of the action potential, $\rm{K^+}$ channels are activated, and more $\rm{K^+}$ ions enter the ECS, raising the Nernst potential $E_{gl,k}^{ex}$ and driving $\rm{K^+}$ into the glial compartment (see Fig. \ref{fig:PotentialFlux4}d). However, if the conductance drops too low (below $\frac{1}{30}$ of the original value), less potassium flows from the ECS into the glial compartment (see Fig. \ref{fig:PotentialFlux4}d), resulting in a lower extracellular potential. Due to the rapid propagation of the electric potential, the non-stimulated region also experiences a drop in extracellular potential (see Fig. \ref{fig:PotentialFlux4}b), which causes the axon membrane potential to increase. This leads to the activation of $\rm{Na^+}$ channels in the non-stimulated region, generating an action potential (see Fig. \ref{fig:MembranePtential4} c-d) and releasing more potassium into the ECS, creating a feedback loop that exacerbates potassium accumulation. In this case, the concentration difference between the stimulated and non-stimulated regions is smaller, and the diffusion flux in the ECS decreases, as shown in Fig. \ref{fig:EcsInnerRIonFlux4}. This is consistent with the clinic's observation that When glial ionic conductance is impaired, it can lead to conditions such as epilepsy, where excessive and synchronized neuronal activity occurs\cite{patel2019neuron}.

 In summary, glial cell ionic conductivity, particularly involving potassium regulation, is crucial for preventing neuronal hyperexcitability and the development of epilepsy. Dysfunction in these mechanisms can lead to disrupted ionic homeostasis, contributing to the occurrence of seizures.

\subsubsection{Connexin Connectivity}

Connexins in glial cells, particularly astrocytes, form gap junctions that facilitate direct communication between cells by allowing the passage of ions, small molecules, and signaling substances. These connexins play a crucial role in maintaining ionic balance, including potassium clearance in the brain's ECS. However, in certain pathological conditions or through experimental interventions, connexin function can be disrupted or blocked \cite{oksanen2019astrocyte}. Such dysfunction reduces the connectivity of the glial compartment. In our simulation, we decrease the diffusion coefficients $D_{gl}^i$ and permeability $\kappa_{gl}$ by factors of $10^{-1}$, $10^{-2}$, and $10^{-4}$ to examine the effects of impaired connexin connectivity.

\begin{figure}
    	\centering
    	\includegraphics[width=1\linewidth]{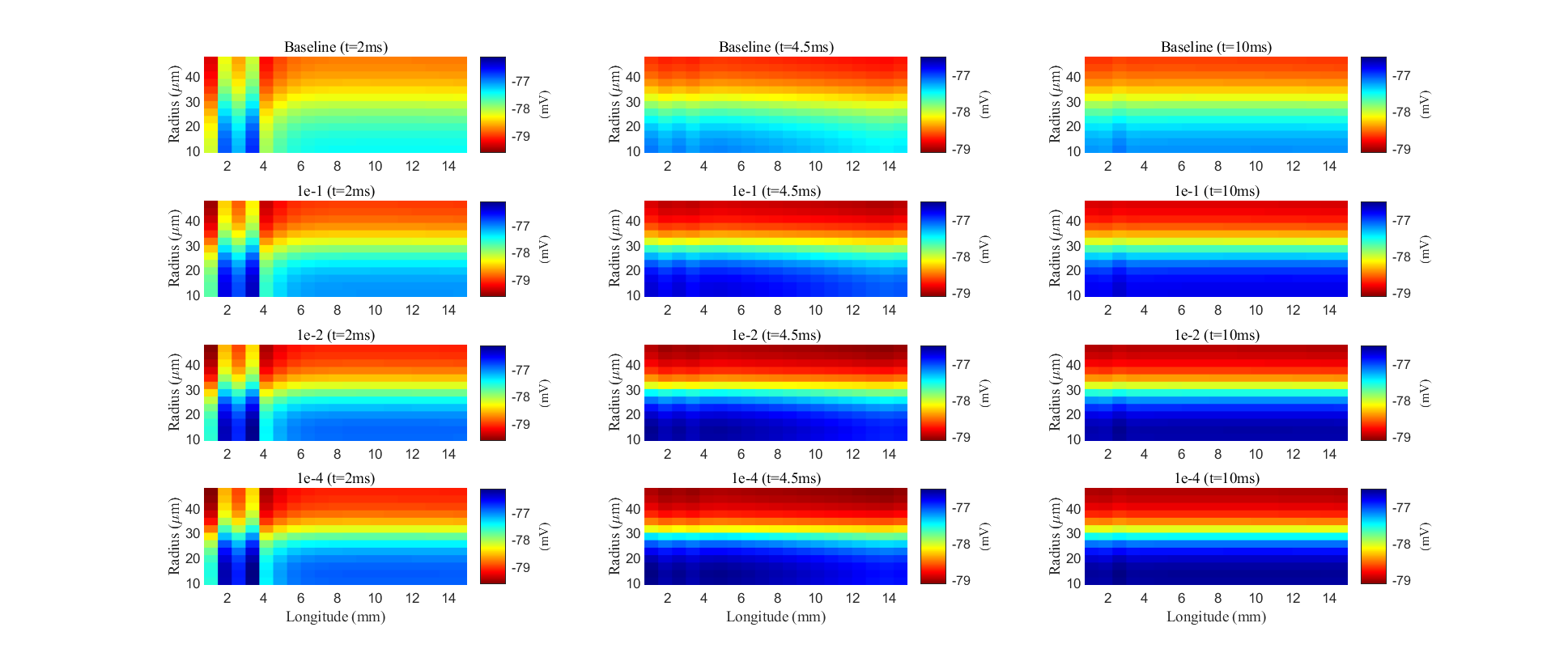}
    	\caption{\label{fig:MemPotential41ms} Spatial distribution of membrane potential during a stimulus in the glial compartment. Different rows are results with different connectivity of glial compartments; Different columns are results at different time slots.}
    \end{figure}

After a stimulus, potassium, and fluid flow from the extracellular space into the glial compartment due to differences in Nernst potential and osmotic pressure. Figure \ref{fig:MemPotential41ms} shows the spatial distribution of electric potential within the glial compartment (long-term behavior can be found in Appendix Fig. \ref{fig:MemPotential41}). As connexin connectivity decreases, the intracellular conductance $\sum_i z_i^2eD_{gl}$ also decreases, which inhibits the propagation of the electric potential from the stimulated region to the non-stimulated region. As a result, the membrane potential in the glial compartment of the stimulated region becomes more elevated compared to the baseline.

\begin{figure}
    	\centering
    	\includegraphics[width=1\linewidth]{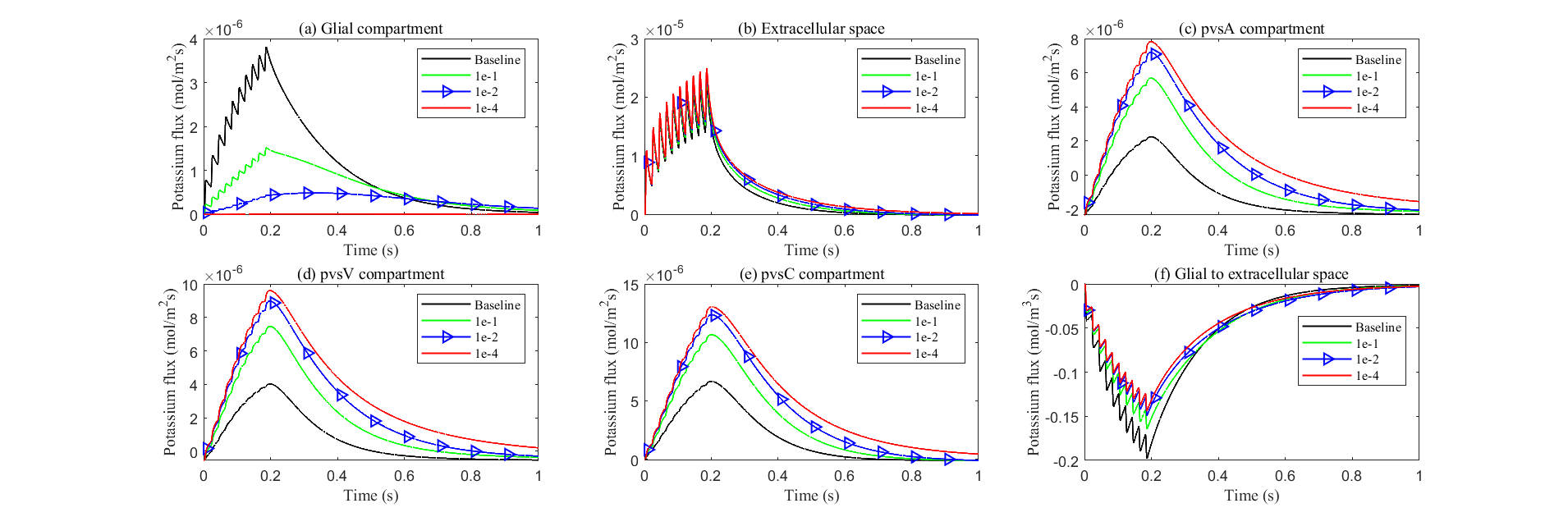}
    	\caption{\label{fig:RfluxIon41} a-e: Average radial potassium flux in the intradomain with varying levels of glial connexin connectivity. f: Average transmembrane potassium flux in the stimulated region.}
    \end{figure}

\begin{figure}
    	\centering
    	\includegraphics[width=1\linewidth]{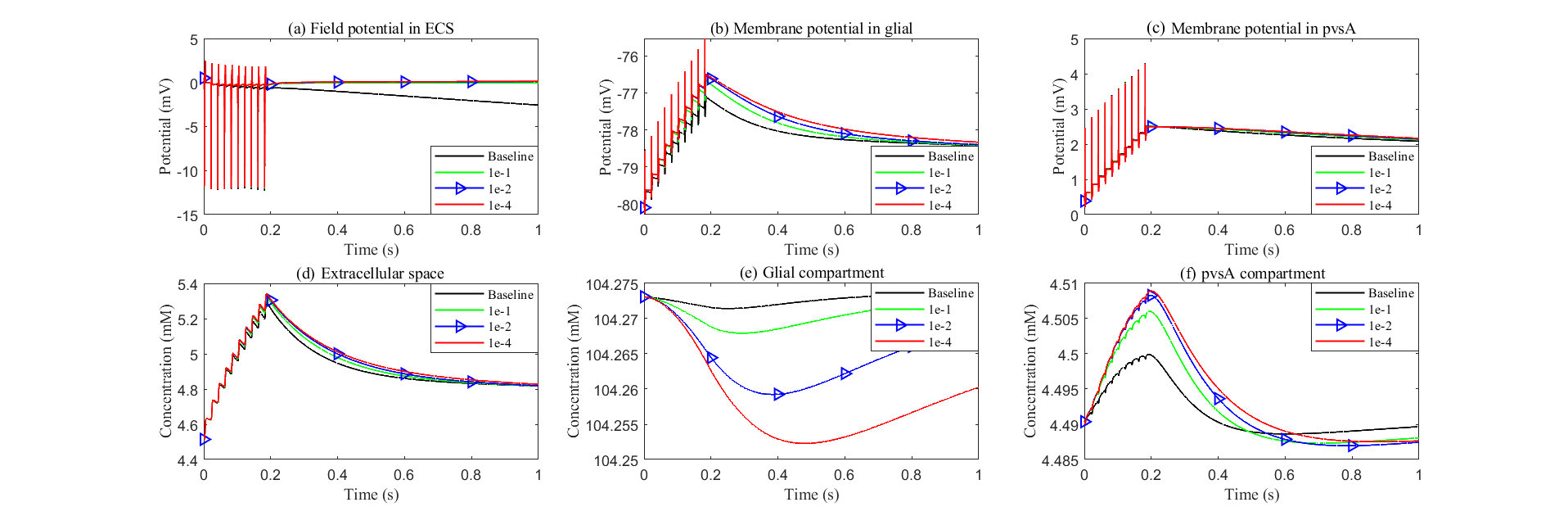}
    	\caption{\label{fig:PhiCon41} a-c: Field potential and membrane potential in the stimulated region. d-f: Potassium concentration in the stimulated region.}
    \end{figure}

Although potassium concentration remains nearly constant (see Fig. \ref{fig:PhiCon41}), due to the swelling of the glial compartment, the rising glial membrane potential (see Fig. \ref{fig:PhiCon41}b) reduces the transmembrane potassium influx, as shown in Fig. \ref{fig:RfluxIon41}f. The reduction in permeability and diffusion within the glial compartment leads to a significant decrease in the potassium buffering flux (see Fig. \ref{fig:RfluxIon41}a). In this scenario, the glial compartment in the stimulated region acts as a reservoir, absorbing excess potassium and protecting the axonal cells.

\begin{figure}
    	\centering
    	\includegraphics[width=1\linewidth]{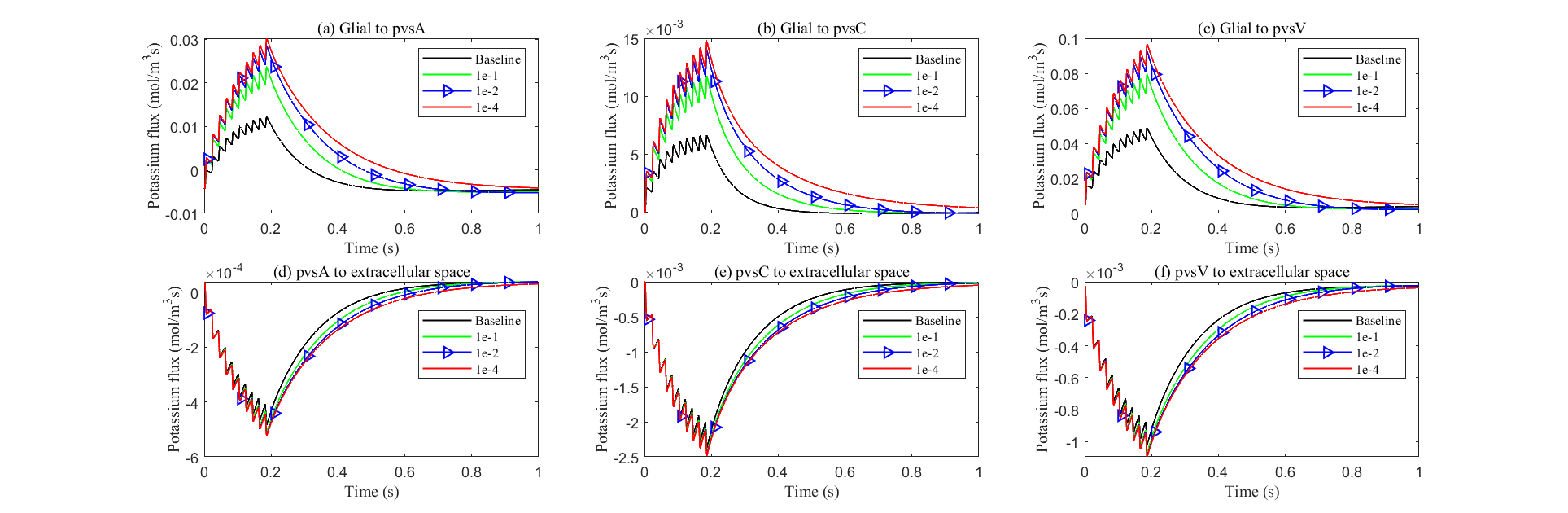}
    	\caption{\label{fig:KConTrans41sti}  Average transmembrane potassium flux  on perivascular spaces with varying levels of glial connexin cnnectivity in the stimulated region.}
    \end{figure}

At the same time, the elevated electric potential in the glial compartment enhances the flux from the glial compartment to the perivascular spaces (see Fig. \ref{fig:KConTrans41sti}a-c), increasing the intra-compartment flux within the perivascular space (see Fig. \ref{fig:RfluxIon41}c-e). This suggests that the perivascular spaces take on a more prominent role in buffering potassium in these conditions (see Fig. \ref{fig:RIonFluxCum41} and Fig. \ref{fig:TransIonCum41} in Appendix). However, compared to the glial electric drift mechanism, convection-based transport in the perivascular spaces is less efficient for potassium clearance (see Fig. \ref{fig:PhiCon41}d).

    In summary, the glial pathway is the most important mechanism for potassium buffering and clearance, while the perivascular space pathway serves as the second most important. When the pathways of glial compartment are compromised, whether it is the glial membrane or the internal pathways of the glial compartment, the perivascular space pathway takes on a more significant role in buffering potassium ions from the ECS.

\subsection{metabolite Clearance}

	This section considers the microcirculation of some uncharged neutral particles, such as nutritional substances or metabolites, such as urea, glutamine, and Creatinine. Then, the transmembrane metabolite flux $J_{l,k}^{Me}$  is in the following form due to the neutrality 
\begin{eqnarray} 
    J_{kex}^{Me} =&C_{up,wind}^{Me}U_{k,ex}- G_k^{Me}  log\left(\frac{C_{ex}^{Me}}{C_{k}^{Me}}\right), k = pa,pv,pc\nonumber\\
    J_{gl,k}^{Me} =&- G_{gl}^{Me}  log\left(\frac{C_{k}^{Me}}{C_{gl}^{Me}}\right),k = ex, pa,pv,pc\nonumber\\
   J_{ax,ex}^{Me} =&- G_{ax}^{Me}  log\left(\frac{C_{ex}^{Me}}{C_{ax}^{Me}}\right),\nonumber 
\end{eqnarray}	
 where $G_{l}^{Me}$ is the transmembrane conductivity of the metabolite. 
	 
	For the boundary conditions in each compartment, we apply membrane boundary conditions at location $\Gamma_2$, homogeneous Neumann boundary conditions at $\Gamma_1$, and Dirichlet boundary conditions at $\Gamma_6$. We implement a no-flux boundary condition at $\Gamma_7$ for the glial compartment. The boundary conditions for the ECS  and the perivascular spaces at $\Gamma_7$ are analogous to those used for ions.
	
	\begin{equation}
		\begin{cases}
			\nabla C_{l}^{Me}\cdot\hat{\mathbf{n}}_{r}=0,\ l=gl,ex,pa,pc,pv & \mbox{on} \ \Gamma_1,\\
			\nabla C_{l}^{Me}\cdot\hat{\mathbf{n}}_{z}=\lambda_{l,left}(C_{l}^{Me}-C_{l}^{Me,\infty}),\ l=ax,gl,ex,pa,pc,pv & \mbox{on} \ \Gamma_2,\\
			\nabla C_{csf}^{Me}\cdot\hat{\mathbf{n}}_{z}=\lambda_{csf,right}(C_{csf}^{Me}-C_{csf}^{Me,\infty}), & \mbox{on} \ \Gamma_3,\\
			C_{l}^{Me}=C_{l}^{Me,re},\ l=ax,gl,ex,pa,pc,pv & \mbox{on} \ \Gamma_6,\\
			\mathbf{j}_{gl}^{Me}\cdot\hat{\mathbf{n}}_{r}=0, & \mbox{on} \ \Gamma_7,\\
			\mathbf{j}_{pa}^{Me}\cdot\hat{\mathbf{n}}_{r}=- G_{pia}^{Me} log\left(\dfrac{C_{csf}^{Me}}{C_{pa}^{Me}}\right)+C_{pa}^{Me}u_{pa,csf}, & \mbox{on} \ \Gamma_7,\\
			\mathbf{j}_{pv}^{Me}\cdot\hat{\mathbf{n}}_{r}=- G_{pia}^{Me} log\left(\dfrac{C_{csf}^{Me}}{C_{pv}^{i}}\right)+C_{pv}^{Me}u_{pv,csf}, & \mbox{on} \ \Gamma_7,\\
			\mathbf{j}_{pc}^{Me}\cdot\hat{\mathbf{n}}_{r}=- G_{pia}^{Me} log\left(\dfrac{C_{csf}^{Me}}{C_{pc}^{Me}}\right), & \mbox{on} \ \Gamma_7,\\
			\mathbf{j}_{ex}^{Me}\cdot\hat{\mathbf{n}}_{r}=- G_{pia}^{Me} log\left(\dfrac{C_{csf}^{Me}}{C_{ex}^{Me}}\right), & \mbox{on} \ \Gamma_7,\\
			\mathbf{j}_{csf}^{Me}\cdot\hat{\mathbf{n}}_{r}=(\mathbf{j}_{pa}^{Me}+\mathbf{j}_{pv}^{Me}+\mathbf{j}_{pc}^{Me}+\mathbf{j}_{ex}^{Me})\cdot\hat{\mathbf{n}}_{r}& \mbox{on} \ \Gamma_7.
		\end{cases}
	\end{equation}
	where $C_{l}^{Me,\infty}$ is the concentration of the metabolic substances in the resting state of each compartment.
	
\begin{figure}
		\centering
		\includegraphics[width=0.8\linewidth]{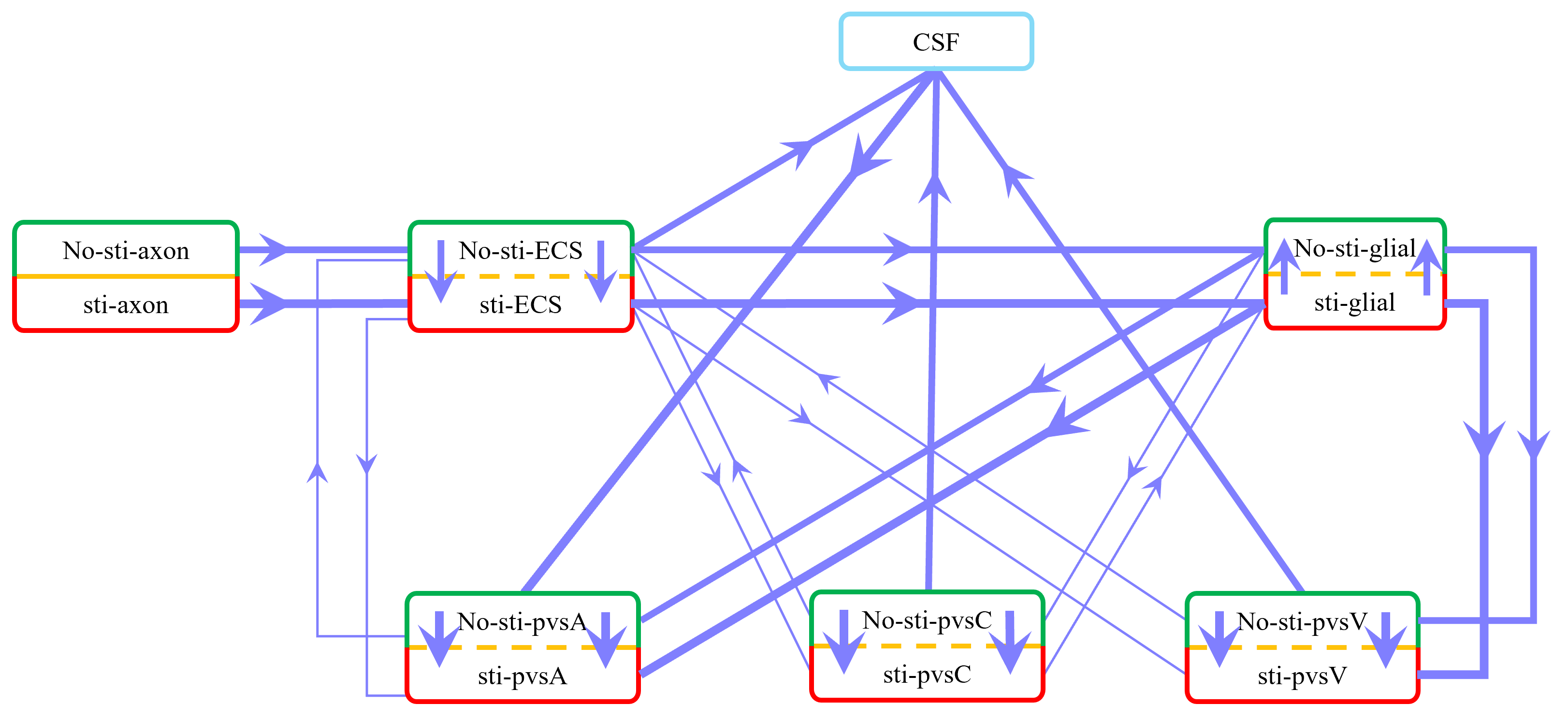}
		\caption{\label{fig:StiMetbolic} Schematic of metabolite flux between the stimulated (lower) and non-stimulated (upper) regions, as well as the transmembrane flux between different compartments during stimulation. Red boxes represent the stimulated regions, and green boxes represent the non-stimulated regions. The thickest lines indicate fluxes around $\rm{10^{-4} ~mol/(m^{3}s)}$, moderately thick lines represent fluxes around $\rm{10^{-5} ~mol/(m^{3}s)}$, and the thinnest lines indicate fluxes less than $\rm{10^{-6} ~mol/(m^{3}s)}$.}
	\end{figure}

    Figure \ref{fig:StiMetbolic} illustrates the microcirculation and clearance pathways of metabolites in the optic nerve. Metabolites enter the optic nerve from the left boundary via the extracellular space (ECS), the glial compartment, and the perivascular space surrounding the vein (pvsV), as well as from the upper boundary through the perivascular space surrounding the artery (pvsA). Once inside the optic nerve, metabolites in both the stimulated and non-stimulated regions move from the axon compartment into the ECS. Within the ECS, metabolites are buffered by the glial compartment and perivascular spaces. After entering the glial compartment, metabolites flow toward either the pvsA or pvsV. Due to the fast fluid flow in the perivascular spaces surrounding the central retinal artery (pvsA) and central retinal vein (pvsV), these spaces play a key role in draining metabolites from the optic nerve. Efficient clearance through these pathways is critical for maintaining optic nerve health and preventing neurodegenerative conditions. This circulation pattern aligns with experimental findings from previous studies \cite{mathieu2017evidence,mogensen2021glymphatic}.
    
\begin{figure}
		\centering
		\includegraphics[width=1\linewidth]{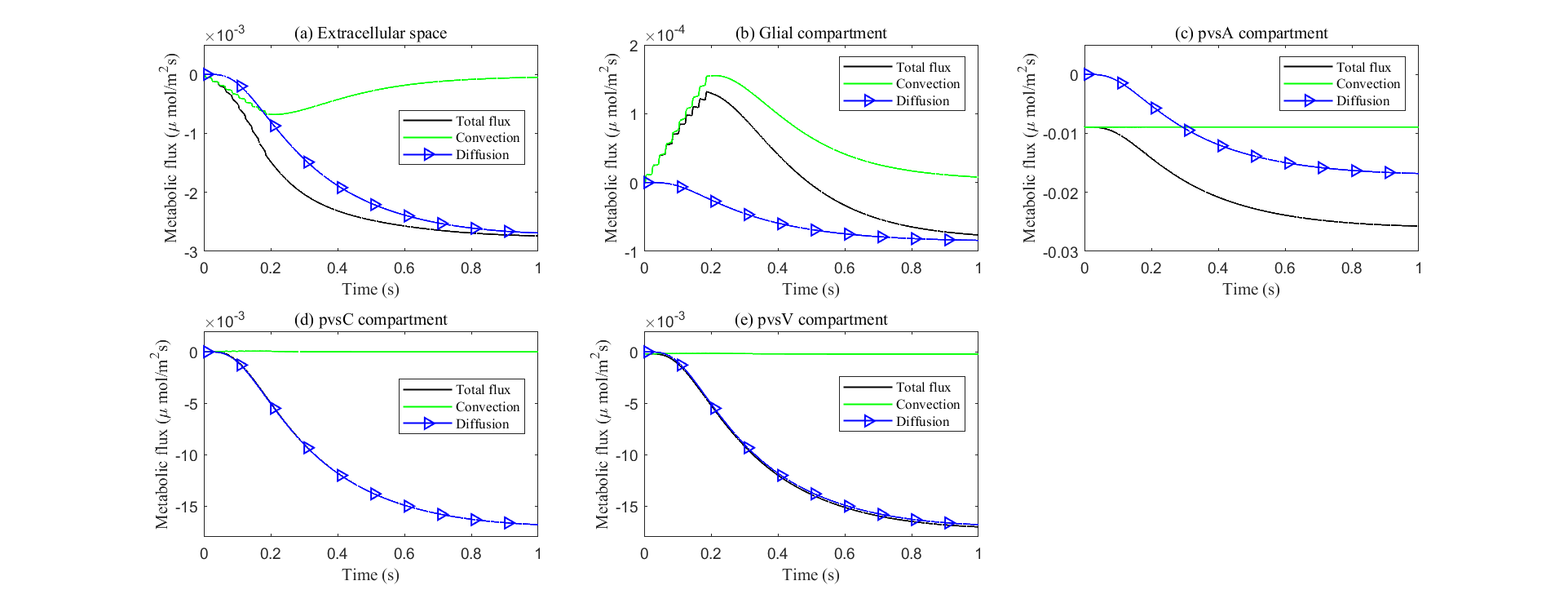}
		\caption{\label{fig:RaStiMet} The radial average metabolite flux in the intradomain.}
	\end{figure}

    Additionally, due to the metabolite drainage by pvsA and pvsV at the lower boundary $\Gamma_1$, the concentration of metabolites in the lower region is lower than in the upper region, making diffusion from the upper region to the lower region the dominant flux mechanism in perivascular spaces and ECS (see Fig. \ref{fig:RaStiMet}). Due to the fluid circulation, the convection flux is the main mechanism for metabolite clearance in the glial compartment. 
 
\begin{figure}
    	\centering
    	\includegraphics[width=1\linewidth]{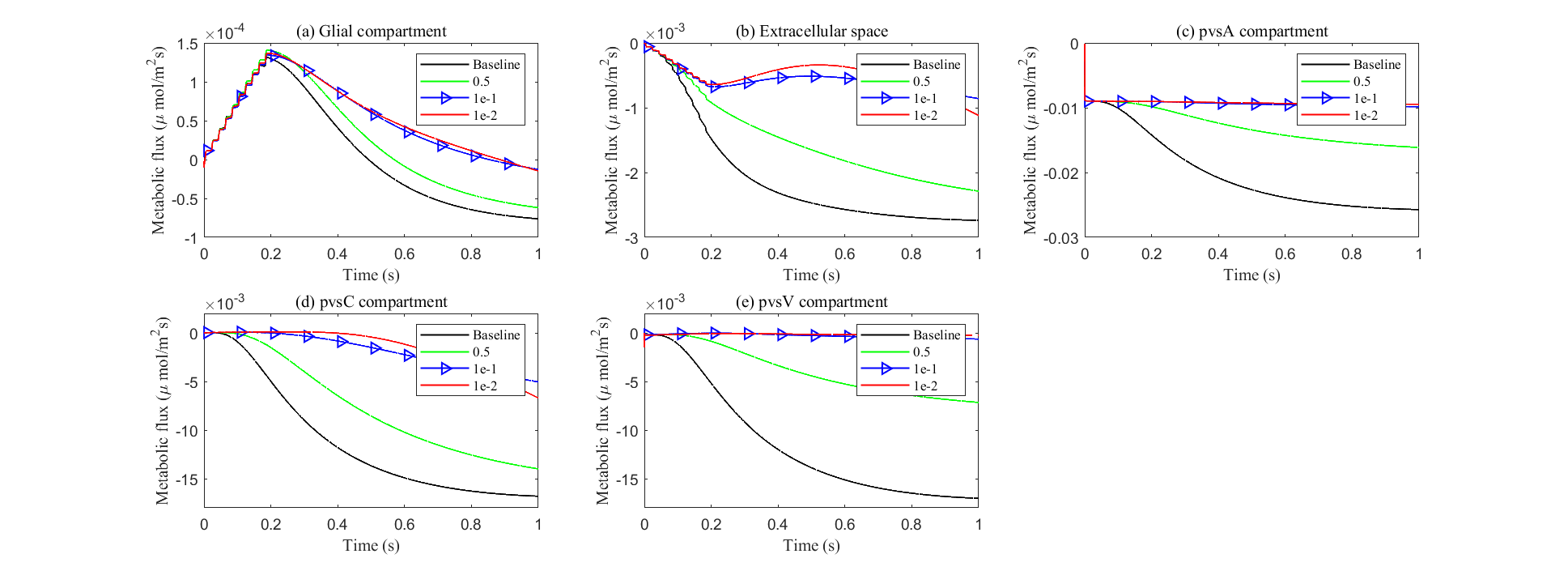}
    	\caption{\label{fig:RaStiMet42} Average radial metabolite flux in the intradomain with varying levels of the permeability within the pvsA and pvsV.}
    \end{figure}

\begin{figure}
    	\centering
    	\includegraphics[width=1\linewidth]{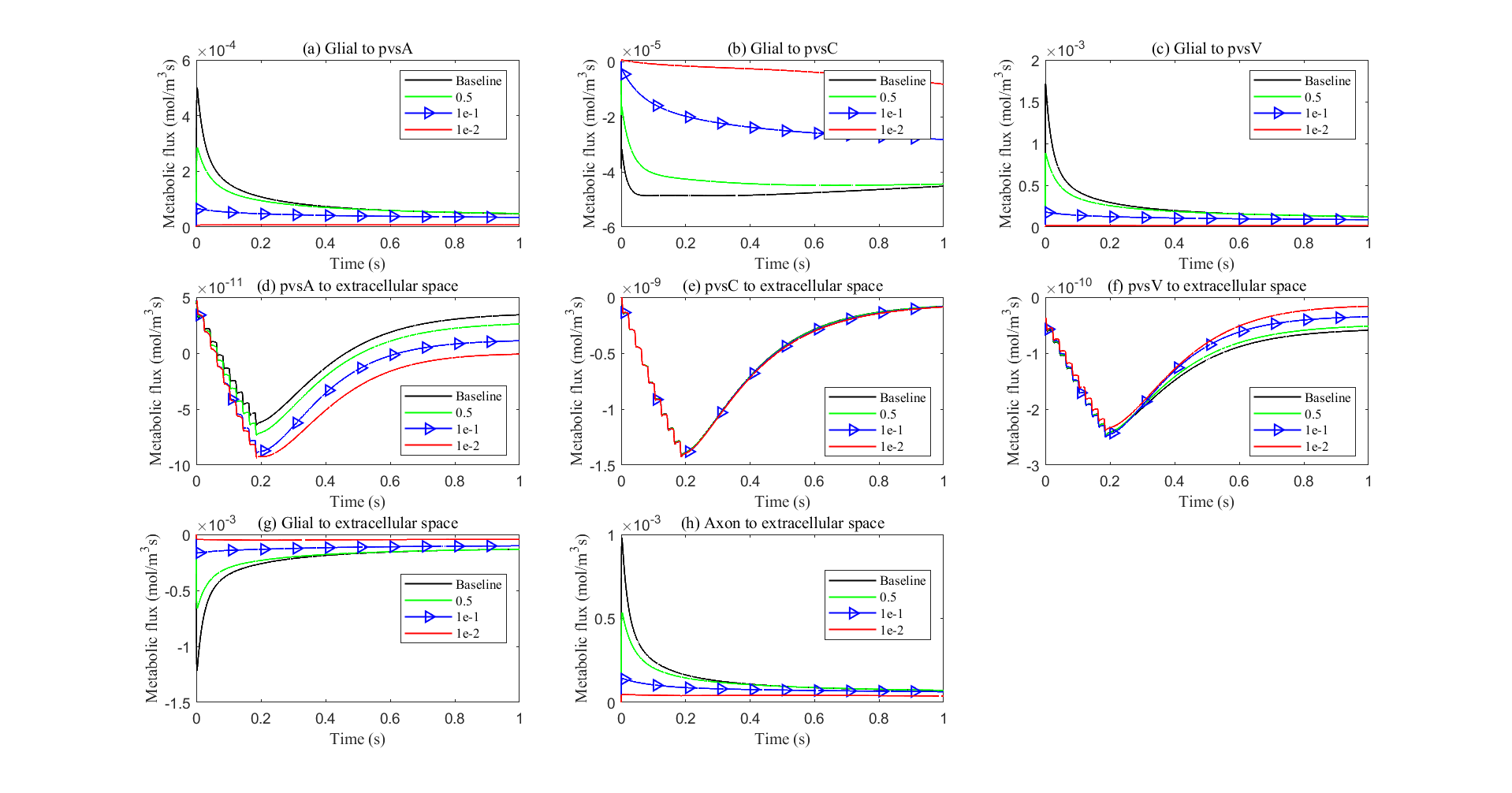}
    	\caption{\label{fig:StiMet42} Average transmembrane metabolite fluxes with varying levels of permeability within the pvsA and pvsV.}
    \end{figure}

\begin{figure}
		\centering
		\includegraphics[width=1\linewidth]{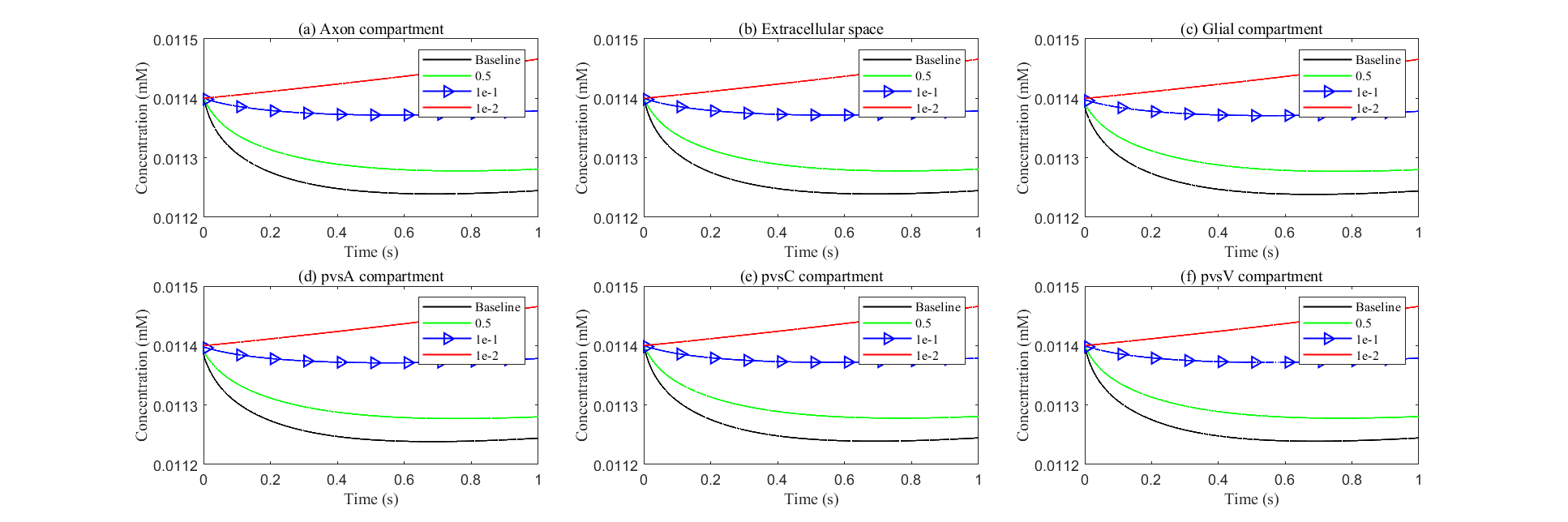}
		\caption{\label{fig:ConMetSti42} Average metabolic waste concentration within each compartment with varying levels of the permeability within the pvsA and pvsV.}
   \end{figure}

Neurodegenerative diseases can also lead to blockages in the perivascular spaces. For instance, in Alzheimer's disease, the accumulation of amyloid-beta in and around blood vessels can obstruct perivascular spaces, impairing the glymphatic system’s ability to clear toxic proteins and waste products. To explore the impact of perivascular spaces on metabolite clearance, we decreased the intradomain diffusion coefficients $D_{l}$ and permeability $\kappa_{l}$ ($l=pa,pv$) by factors of $0.5$, $10^{-1}$, and $10^{-2}$. As the permeability of pvsA and pvsV decreases, the radial total flux between the stimulated and non-stimulated regions is significantly reduced (see Fig. \ref{fig:RaStiMet42}). This reduction impairs the drainage of metabolites, leading to their accumulation in pvsA and pvsV (see Fig. \ref{fig:ConMetSti42}d$\&$f). The buildup of metabolites subsequently reduces the transmembrane flux from the glial compartment to pvsA/V (see Fig. \ref{fig:StiMet42}), which is the primary pathway for metabolite clearance. Consequently, metabolic waste accumulates within each compartment (see Fig. \ref{fig:ConMetSti42}).

In summary, the perivascular spaces play a critical role in metabolite clearance. When these spaces are compromised, metabolites cannot be cleared in a timely and efficient manner, leading to their accumulation and potentially exacerbating neurodegenerative conditions.

\section{Conclusions and Discussion}

The central insight of this study is that "everything influences everything else" in complex biological systems. Fluid flow in biological contexts is driven by a combination of convection (influenced by hydrostatic and osmotic pressure ), diffusion (influenced by concentration gradients), and migration (driven by electric fields). These mechanisms are highly interconnected and cannot be considered in isolation. Biological responses to parameter changes are often unpredictable due to the complex interactions between these forces, which may shift in significance depending on the physiological or pathological context.

Our study emphasizes the necessity of applying complex fluid theory to analyze the microcirculation and metabolite clearance mechanisms in the optic nerve, expanding upon the pioneering work of Richard Orkand and the Harvard group \cite{1966Effect}. We introduced a multicompartment model that incorporates the axon,  glial, ECS compartment,  pvsA, pvsV, and pvsC to provide a comprehensive framework for analyzing these complex interactions, particularly focusing on the role of glial cells and perivascular spaces in potassium and metabolite clearance.

Our simulations provide an in-depth understanding of how glial cells and perivascular spaces contribute to potassium buffering and clearance in the narrow extracellular space. Glial cells play a crucial role in absorbing excess potassium from the ECS during neuronal firing, while perivascular spaces assist by providing an additional pathway for clearance. Potassium flux moves from the axon into the ECS during stimulation, accumulates, and is then buffered through both the glial compartment and perivascular spaces. This coordinated process ensures potassium is transferred from the stimulated region to non-stimulated areas, helping to maintain ionic homeostasis during neural activity.

The integration of convection, diffusion, and electrical migration in our model underscores the complex nature of ion movement in the optic nerve. Potassium clearance is primarily driven by electric drift within glial syncytia, which enables efficient redistribution and elimination of excess potassium. These findings reinforce the importance of glial cells' structural and functional properties, particularly their role in facilitating convective flow through interconnected glial syncytia. Perivascular spaces, particularly those surrounding the central retinal artery and vein, play a vital role in fluid circulation and potassium clearance, further enhancing the buffering capacity of the optic nerve.

Our study also extends to the clearance of metabolites, highlighting the role of perivascular spaces in this process. Blockage of perivascular spaces, as seen in neurodegenerative diseases like Alzheimer's, can impair the glymphatic system’s ability to clear toxic proteins and waste products such as amyloid-beta. By simulating reduced permeability and diffusion in perivascular spaces, we demonstrated how compromised perivascular pathways lead to metabolite accumulation in both the pvsA and pvsV compartments. This buildup of metabolic waste reduces the efficiency of transmembrane flux from the glial compartment, leading to impaired clearance of metabolites throughout the optic nerve. These results align with experimental findings on the importance of perivascular spaces in maintaining neuronal health and underscore the risks associated with their dysfunction.

The findings of this study have several important applications and pave the way for future research. First, the model can be extended to simulate disease conditions, such as glaucoma or Alzheimer’s disease, where impaired fluid dynamics and clearance mechanisms play a critical role. The insights gained from this model could also help in developing therapeutic strategies that target glial cells or perivascular spaces to enhance metabolite clearance and maintain neuronal health.

Future work could involve refining the model to explore how various pathological conditions, such as oxidative stress, inflammation, or amyloid-beta accumulation, affect perivascular space function and glial cell connectivity. Additionally, incorporating more detailed biochemical interactions, such as enzyme activity in metabolite degradation, could enhance our understanding of waste clearance processes. Machine learning approaches could be applied to optimize parameter selection and explore the nonlinear relationships between fluid flow, ion transport, and metabolite clearance under varying physiological conditions.

\section*{Acknowledgments}
This work was partially supported by the 
National Natural Science Foundation of China no. 12231004 (H. Huang) and 12071190 (S. Xu).

\bibliographystyle{unsrt}
\bibliography{sample}

\newpage
\appendix

	\begin{figure}
		\centering
		\includegraphics[width=0.8\linewidth]{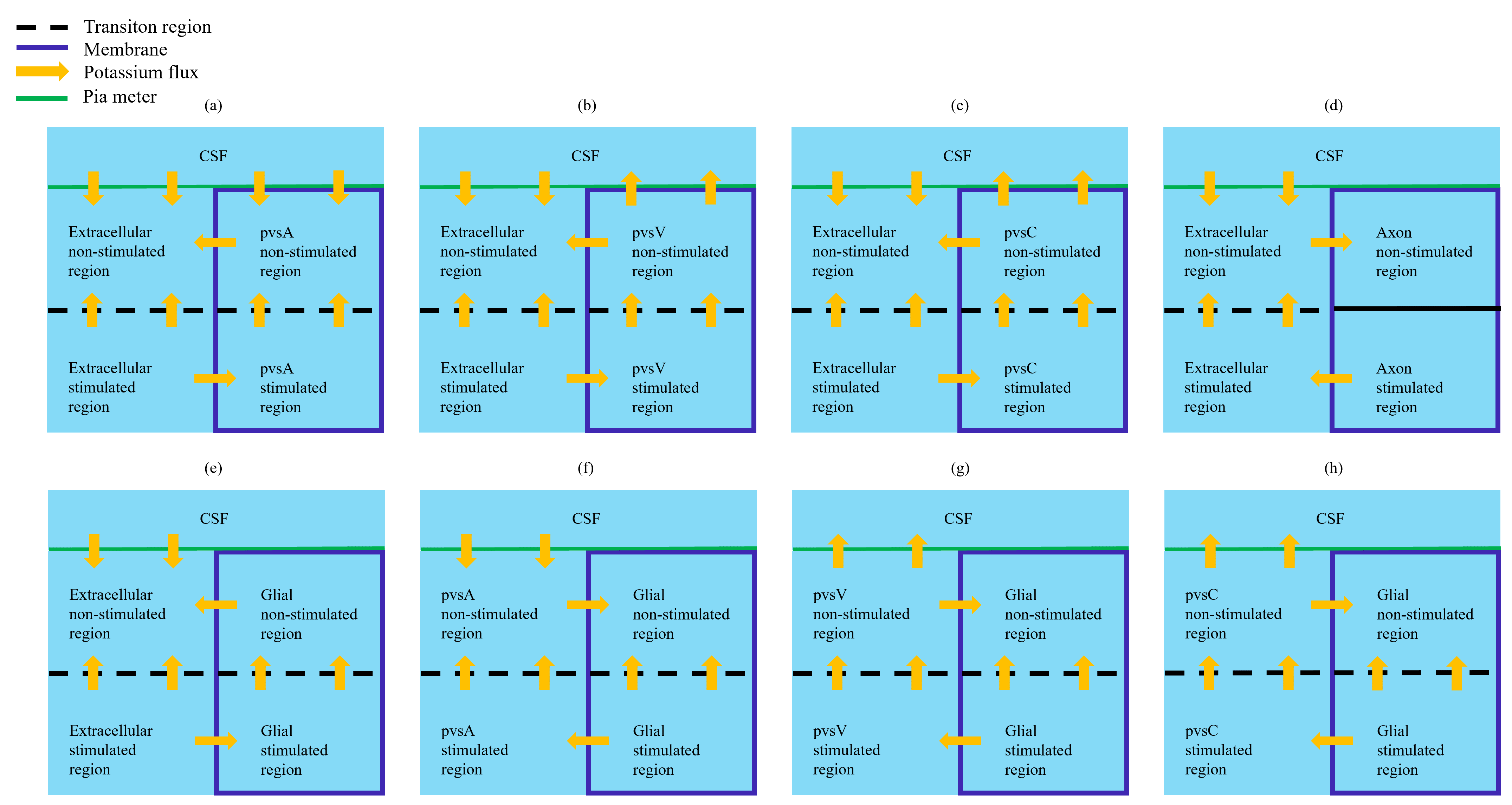}
		\caption{\label{fig:IonTransition} Schematic graph of the potassium microcirculation between two compartments when inner part axon was stimulated.   (a) Extracelluar-pvsA; (b) Extracelluar-pvsV; (c) Extracelluar-pvsC; (d) Extracelluar-Axon; (e) Extracellular-Glial; (f) Glial-pvsA; (g) Glial-pvsV; (h) Glial-pvsC; The dash black line is the interface between stimulus region and unstimulus region.   The solid black line in axon compartment means the potassium flux only in $z$ direction and can not transition between the stimulated and non-stimulated region.}
	\end{figure}

\begin{figure}
    \centering
    \includegraphics[width=0.8\linewidth]{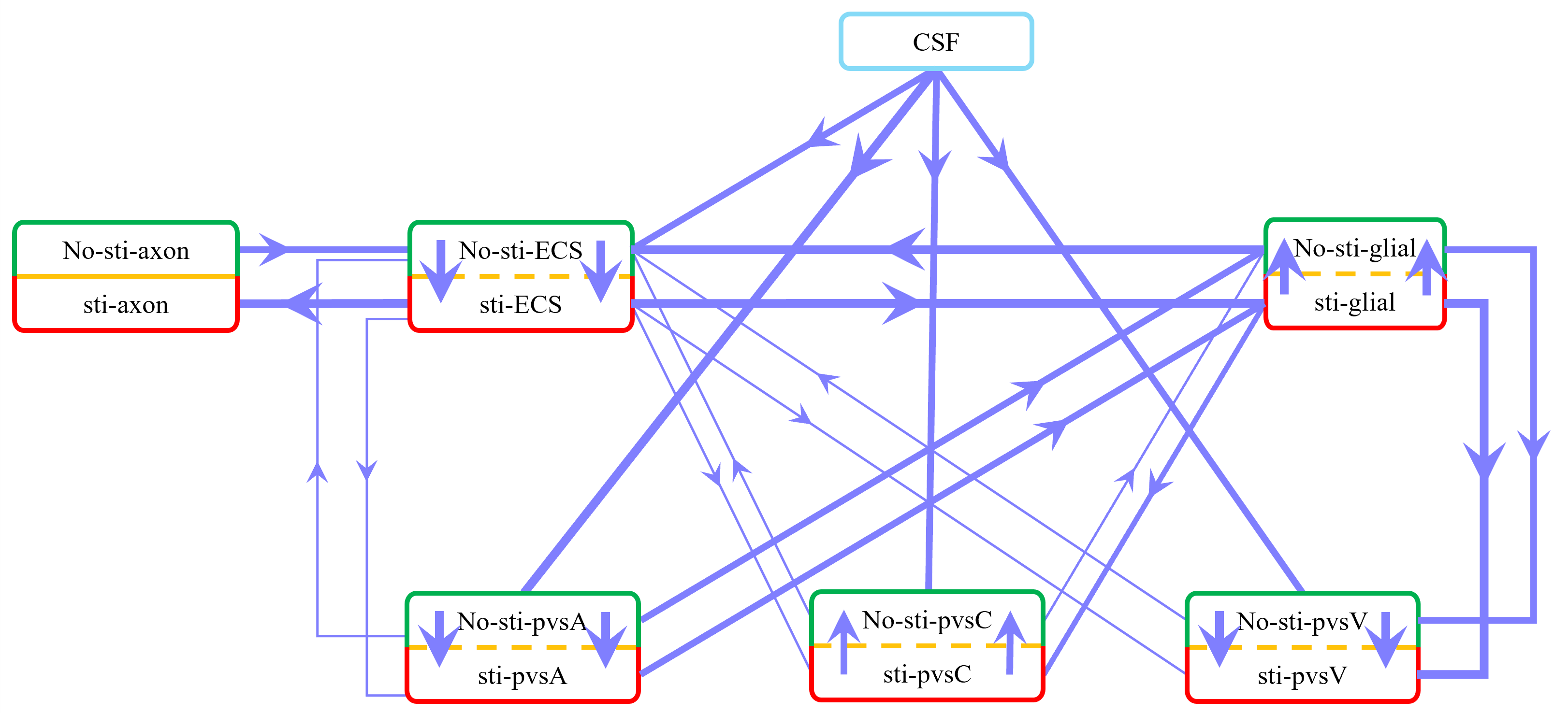}
    \caption{\label{fig:TransFluidFlux} Schematic of fluid flux between the stimulated (lower) and non-stimulated (upper) regions, as well as transmembrane fluid flux between different compartments during stimulation. Red boxes represent the stimulated regions, and green boxes represent the non-stimulated regions. The thickest lines indicate fluxes around $\rm{10^{-3} ~/s}$, moderately thick lines represent fluxes around $\rm{10^{-4} ~/s}$, and the thinnest lines indicate fluxes less than $\rm{10^{-5} ~/s}$. The transmembrane fluid flux is $\mathcal{M}U$.}
\end{figure}
 
	

        \begin{figure}
	       \centering
		\includegraphics[width=1\linewidth]{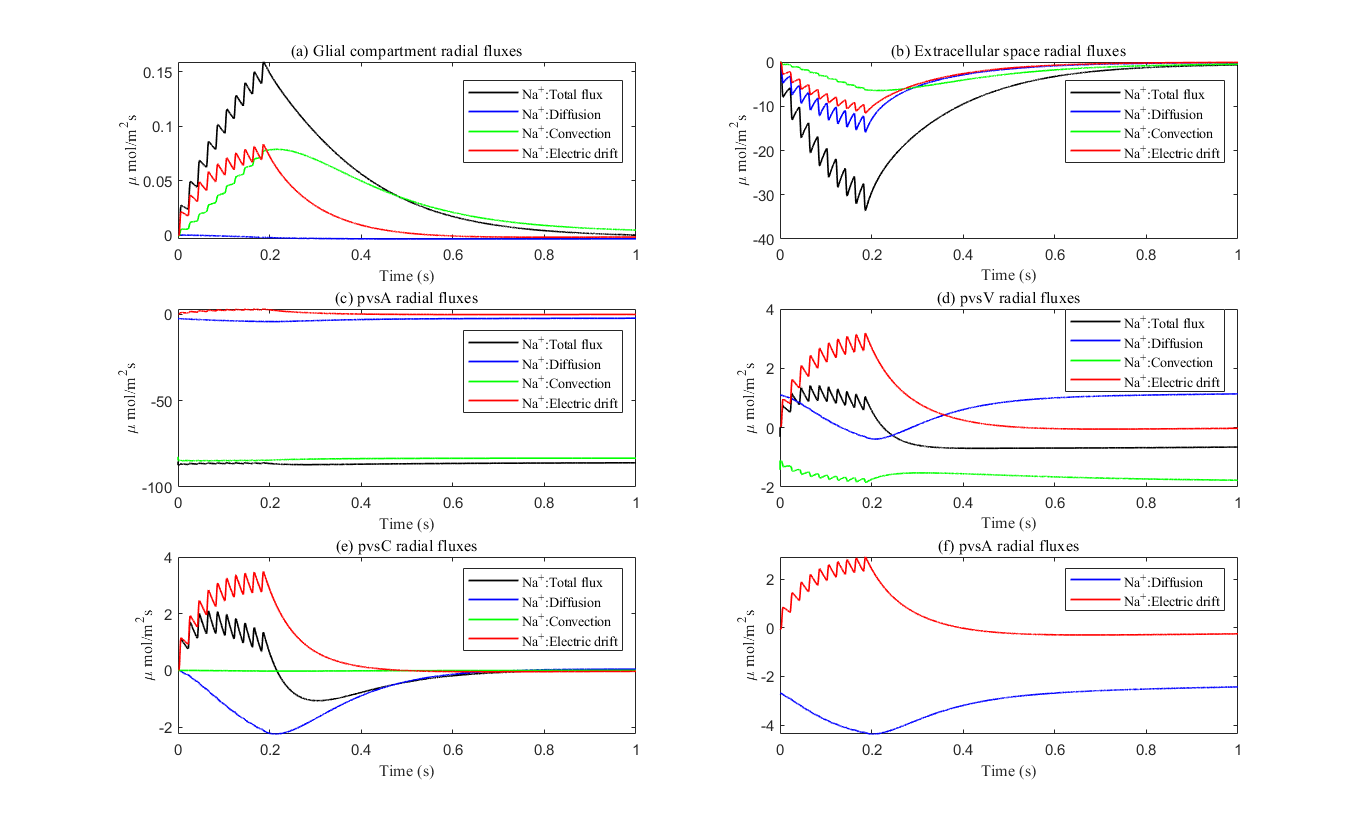}
		\caption{\label{fig:NaDiffusion} Average radial direction $\rm{Na^{+}}$ fluxes components in the intradomain.}
	\end{figure}

    \begin{figure}
    	\centering
    	\includegraphics[width=1\linewidth]{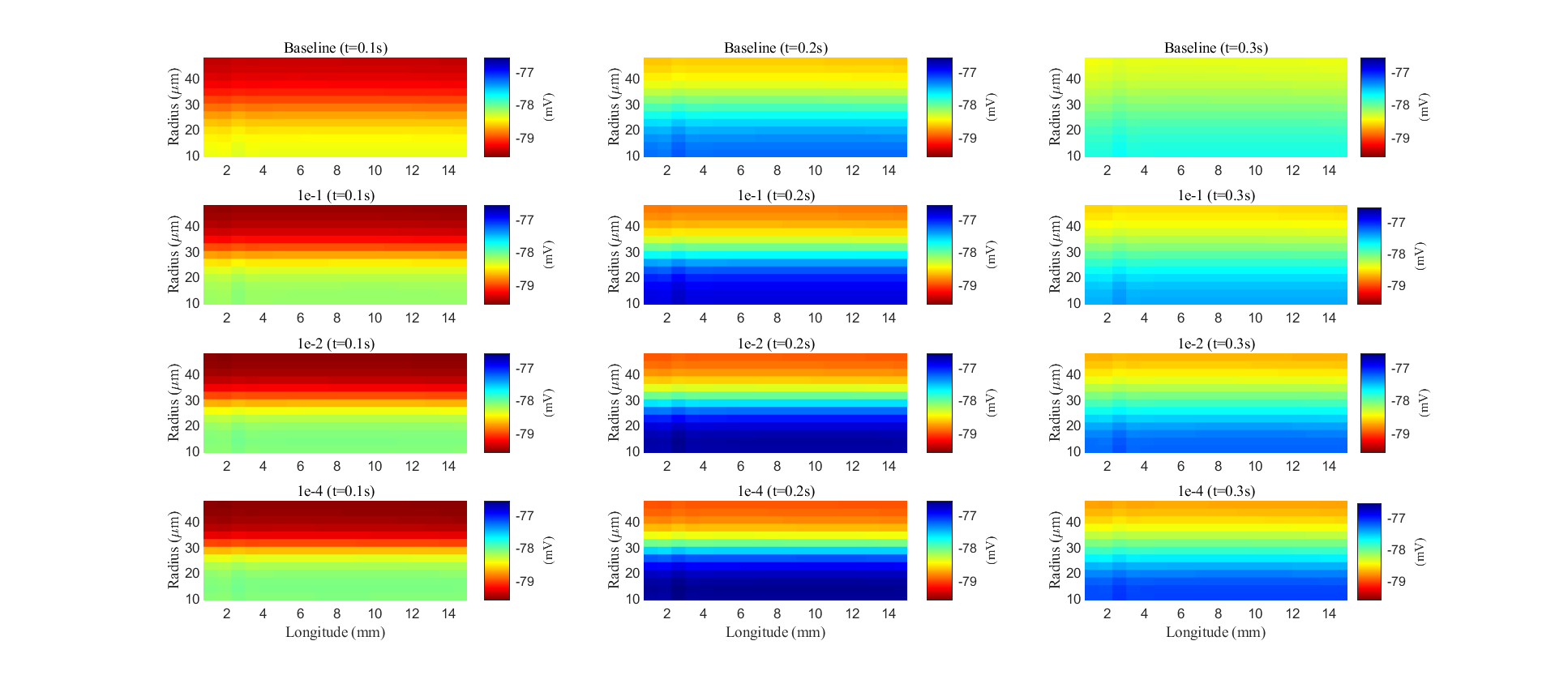}
    	\caption{\label{fig:MemPotential41} Spatial distribution of membrane potential during and after a train of stimuli in the glial compartment. Different rows are results with different connectivity of glial compartment; Different columns are results at different time slots.}
    \end{figure}

 \begin{figure}
    	\centering
    	\includegraphics[width=1\linewidth]{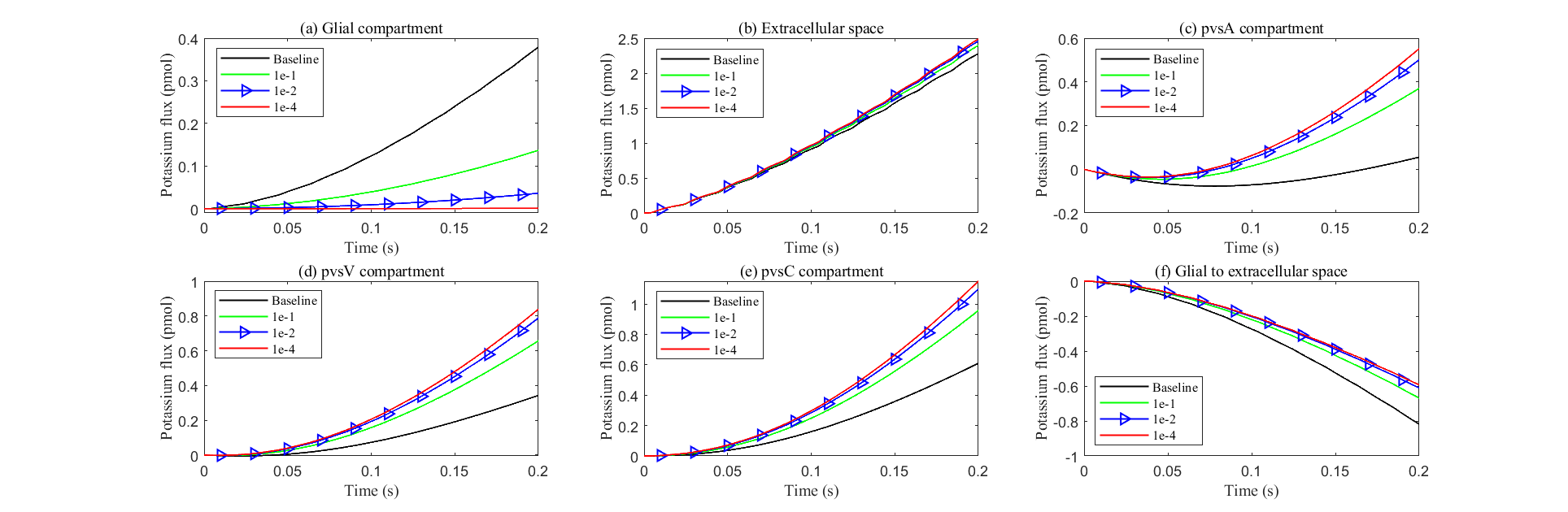}
    	\caption{\label{fig:RIonFluxCum41} Cumulative potassium fluxes during the stimulus with varying levels of glial connexin connectivity. a-e: Radial cumulative potassium fluxes within compartments. f: Cumulative transmembrane potassium fluxes in the stimulated region.}
    \end{figure}

    \begin{figure}
    	\centering
    	\includegraphics[width=1\linewidth]{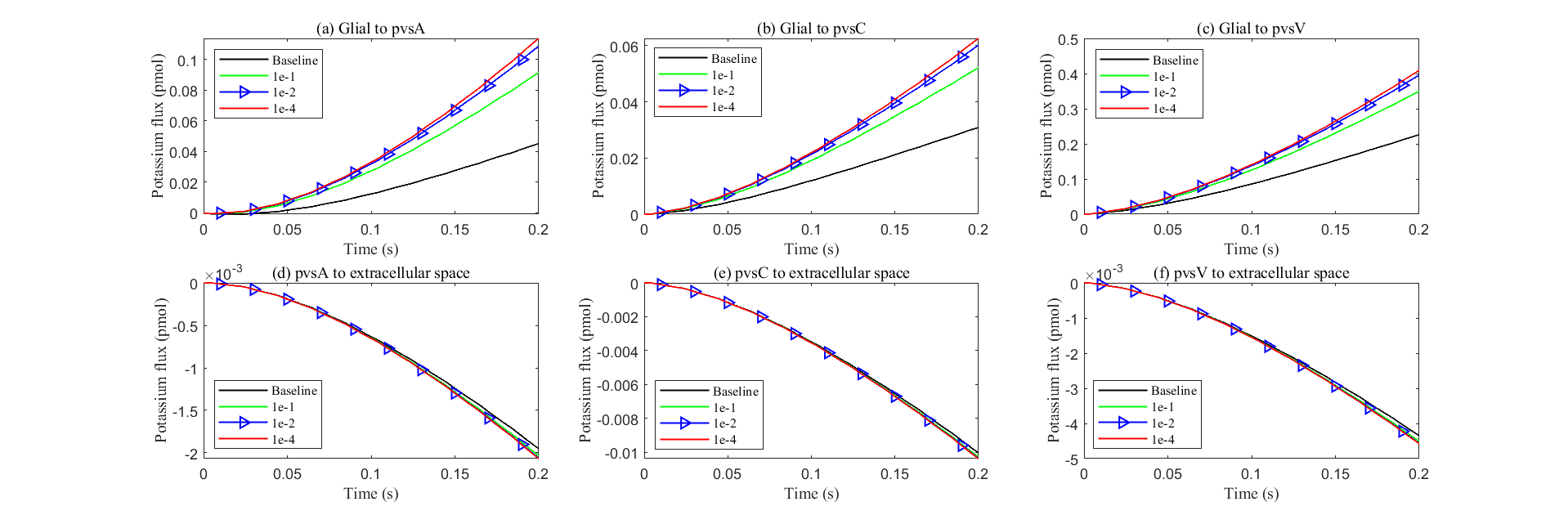}
    	\caption{\label{fig:TransIonCum41} Cumulative transmembrane potassium fluxes in the stimulated region during the stimulus with varying levels of glial connexin connectivity.}
    \end{figure}

\end{document}